\documentclass[12pt,reqno]{amsart}
\usepackage[centertags]{amsmath}
\usepackage{amsfonts}
\usepackage{amssymb}
\usepackage{amsthm}
\usepackage{amsmath}
\usepackage[top=2.5cm, bottom=2.5cm, left=2.5cm, right=2.5cm]{geometry}
\usepackage{color}
\usepackage{graphicx}

\usepackage{hyperref}
\hypersetup{
    colorlinks=true,
    linkcolor=blue,
    filecolor=magenta,
    urlcolor=cyan,
}
\numberwithin{equation}{section}

\renewcommand{\epsilon}{\varepsilon}

\newcommand{\be}{\begin{equation}}
\newcommand{\ee}{\end{equation}}

\newcommand{\cB}{\mathcal{B}}
\newcommand{\C}{\mathbb{C}}
\newcommand{\F}{\mathbb{F}}
\newcommand{\R}{\mathbb{R}}
\newcommand{\T}{\mathbb{T}}

\newcommand{\Z}{\mathbb{Z}}

\newcommand{\N}{\mathbb{N}}
\newcommand{\K}{\mathbb{K}}

\newcommand{\cC}{\mathcal{C}}
\newcommand{\cD}{\mathcal{D}}

\newcommand{\cE}{{\mathcal E}}
\newcommand{\cF}{\mathcal{F}}

\newcommand{\Rho}{\mathcal{R}}
\newcommand{\cQ}{\mathcal{Q}}

\newcommand{\fW}{\mathfrak{W}}
\newcommand{\fS}{\mathfrak{S}}
\newcommand{\fD}{\mathfrak{D}}
\newcommand{\fU}{\mathfrak{U}}

\newcommand{\bH}{{\bf H}}
\newcommand{\bV}{{\bf V}}
\newcommand{\bU}{{\bf U}}

\renewcommand{\i}{\mathrm{i}}
\newcommand{\x}{{\bf x}}
\newcommand{\y}{{\bf y}}
\newcommand{\p}{{\bf p}}
\newcommand{\q}{{\bf q}}
\renewcommand{\k}{{\bf k}}

\newcommand{\supp}{{\ensuremath{\mathrm{supp\,}}}}
\newcommand{\disc}{{\ensuremath{\mathrm{disc}}}}
\newcommand{\ess}{{\ensuremath{\mathrm{ess}}}}

\renewcommand{\epsilon}{\varepsilon}
\newcommand{\qoshuv}{\displaystyle\oplus}

\theoremstyle{definition}
\newtheorem{definition}{Definition}[section]
\newtheorem{remark}{Remark}[section]

\theoremstyle{plain}
{\bf}{\itshape}
\newtheorem{theorem}{Theorem}[section]%
\newtheorem{lemma}{Lemma}[section]
\newtheorem{corollary}{Corollary}[section]

\newtheorem{claim}{Step}
\theoremstyle{definition}

\begin{document}

\title{Existence of bound states of $N$-body problem in 
an optical lattice}

\author{Shokhrukh Yu. Kholmatov$\!^{1,2,3}$, Zahriddin I. Muminov$\!^4$}
\address{$\!^1\!$International Centre for Theoretical Physics (ICTP),
Strada Costiera 11, 34151 Trieste, Italy}
\address{$\!^2\!$Scuola Internazionale Superiore di
Studi Avanzati (SISSA),
via Bonomea 265, 34136 Trieste, Italy}
\address{$\!^3\!$Fakult\"at f\"ur Mathematik Universit\"at Wien,
Oskar-Morgenstern-Platz 1, 1090 Vienna, Austria}
\email{$^{1,2,3}\!$shokhrukh.kholmatov@univie.ac.at}

\address{$\!^4\!$Nilai University, 
1, Persiaran Universiti, Putra Nilai, 
71800 Nilai, Negeri Senbilan, Malaysia}
\email{$\!^4\!$zimuminov@mail.ru}

\begin{abstract}
We provide sufficient conditions to have at least one 
$N$-particle bound state below  the essential spectrum of 
a large class of $N$-particle discrete Schr\"odinger 
operators $H(K),$ $K\in \T^d,$ $d\ge1,$ associated 
to the Hamiltonian of (not necessarily identical) 
$N$ particles, moving  on 
a lattice  $\Z^d$ and interacting via
short-range pair potentials. 
We also describe the essential spectrum of $H(K).$

\end{abstract}


\maketitle

\section*{Introduction}

One of the remarkable results in the spectral theory of 
multi-particle continuous Schr\"odinger operators is the 
description of  essential spectrum (the HVZ theorem in honour 
of W. Hunziker \cite{Hun}, C. van Winter \cite{Win} and 
G. Zhislin \cite{Zhis}): 
{\it the essential spectrum of 
an $N$-particle Hamiltonian (in the center-of-mass frame) 
is a half-line whose lowest bound is the lowest
possible energy which two independent subsystems can have}.
Since then the result has been substantially
improved and extended 
to various classes of operators
(see the survey \cite{HS:2000} and references therein).

Few results are available in the literature  
on  essential spectra of 
 discrete Schr\"odinger operators associated to 
 many-body systems in an optical lattice $\Z^d,$ $d\ge1:$   
the essential spectrum of a three-body problem 
on $\Z^3$ with analytic dispersion functions was described in   
\cite{ALM:07:Nach};  a four-body HVZ theorem with  
discrete Laplacian and zero-range potentials in $\Z^3$  
was proved in
\cite{ALA:2003,Mum}; see also \cite{Mog} and references therein 
for other results related to  spectral properties of 
multi-particle lattice Schr\"odinger operators.

One of the fundamental  differences 
between the multi-particle continuous Hamiltonian in $\R^d,$ $d\ge1,$ 
and the discrete Hamiltonian in 
$\Z^d$  is that the latter 
is not rotationally invariant.
However using a technique of separation of variables --
a lattice analogue of the center-of-mass frame 
\cite{GSch:97,Lak:1993,Yaf:2000}  (see also 
Section \ref{sec:aniqla_dsho} of the present paper),
the discrete Hamiltonian can be decomposed into  fibers, i.e.
it can be represented as a direct integral of 
a family of discrete Schr\"odinger operators $H(K),$ 
parametrized by the so-called 
\emph{$N$-particle quasi-momentum} $K\in \T^d,$ where 
$\T^d$ -- the $d$-dimensional torus \cite{Mat,Mog}. 
In contrast to the continuous case, fibers non-trivially
depend on the quasi-momentum $K,$ and therefore, their spectra are 
quite sensitive to the change  of $K:$ 
even in the two-particle case, the essential spectrum may collapse to
a point, and hence, it is not absolutely continuous \cite[Remark 2.1]{LH:2011}.  
Moreover, by virtue of the  boundedness of $H(K),$ 
its essential spectrum is no longer a half-line on the real axis, but 
an at most countable  union of closed segments (see 
Theorem \ref{teo:HVZ0} below) 
and in turn this may allow the Efimov effect 
to appear not only at the lower edge of the essential spectrum, but also
at the edges of gaps between those segments \cite{Mum9}.

Discrete Schr\"odinger operators in lattice  and 
their applications in solid-state physics were duly stressed, 
for instance, in \cite{GSch:97,Mat,Mog,RSIV}; we also refer  to 
\cite{JBC:1998,LSA:2012,Wall:2015,Wink:2006} and 
references therein for  experimental 
and theoretical results in the theory of ultracold 
atoms on optical lattices.

In the first part  of this paper we consider 
the discrete Schr\"odinger operator $H(K),$ $K\in \T^d,$ 
associated to the Hamiltonian of a system of $N\ge2$ particles moving
on a $d$-dimensional lattice $\Z^d$ and interacting via short-range
pair potentials. We prove an analogue of the HVZ theorem using the
diagrammatic method for a large class of 
potentials and dispersion relations that are not necessarily of
compact support. More precisely, we show that under Hypotheses 
\eqref{shart_dispersion}-\eqref{shart:potential} below, 
{\it the essential spectrum of 
$H(K)$ is a union of spectra of all two-cluster operators and  consists
of an at most countable union of disjoint closed segments; the only 
accumulation points (if any) of 
eigenvalues of $H(K)$ outside the essential spectrum are the edges 
of those segments} (Theorem \ref{teo:HVZ0}).

One of the practical applications of the HVZ theorem is that
it allows to use  variational techniques more efficiently 
to study  eigenvalues and eigenvectors of $H(K).$ Further 
eigenvectors of $H(K)$ will  also be called 
$N$-particle bound states.
There is a considerable literature devoted to the 
finiteness of bound states of  continuous Schr\"odinger operators
with  short-range pair potentials 
(see e.g. \cite{BMO:2018,KS:1980,Sig:1976,Sim,Yo:2017}
and references therein). Nevertheless, apart from the Efimov effect 
(see e.g. \cite{BT:2017,Gr:2014,Sob:1993,Tam:1991,Yaf:1974} and references 
therein), not much seems known on the existence of bound states;
some sufficient conditions to 
have an $N$-particle bound state
can be found, for example, in \cite{ChM:1980,BCK:1986,PMC:1985},
spectral properties 
(including existence and non-existence of discrete spectrum)
of $2+1$ fermionic trimers with contact interactions have been 
studied in \cite{BMO:2018}; see also \cite{Zhis:1960}.

In the discrete case, the Efimov effect was studied, for instance, in
\cite{ALM:04:Puan,Lak:1993}. 
The existence of  three-particle bound states  
of purely attractive 
and purely repulsive systems of three identical bosons, 
interacting via  zero-range pair potentials
in $\Z^d,$ $d=1,2,$  
has been recently established in \cite{LDK:2016}; same results 
still hold for the purely attractive 
or purely repulsive  system of $2+1$ fermionic trimers 
in $\Z^d,$ $d=1,2,$ interacting via zero-range pair potentials \cite{LL:2017}.

In the second part of the present paper, we provide sufficient conditions
to have at least one $N$-particle bound state of a purely 
attractive (resp. purely repulsive) system of particles. 
The main result here is that 
{\it if any two-particle subsystem  has a bound state below (resp. above) 
its essential spectrum, then the  discrete spectrum, 
below (resp. above) the essential spectrum, of 
the $N$-particle Schr\"odinger operator 
is nonempty } 
(Theorems \ref{teo:existence_suff} and \ref{teo:existence_suff_rep}). 
To the best of our knowledge, such a result has not been published yet in 
the continuous case.

An advantage of this result in applications is that 
spectral properties of two-particle 
discrete Schr\"odinger operators in  
lattice have been sufficiently well-studied 
(see e.g. \cite{ALMM:2006,LH:2011} and references therein),
in particular, in $\Z^d,$ $d=1,2,$ a two-particle 
bound state always exists, hence our result generalizes
also the results of \cite{LDK:2016} 
(see Section \ref{sec:examples}).
However, recall that we provide only sufficient conditions:
one counterexample would be the Efimov effect.

The universality of our existence results is 
that we claim the existence of 
bound states in {\it every} purely attractive 
or purely repulsive 
system of arbitrary (finite) number of particles  
as soon as  pairwise interactions 
between particles are strong enough. 
Such  universality is less obvious,
for example, in the Efimov physics (see  \cite[Section 1.2]{NE:2017}),
since currently the Efimov effect is known only for three-particle 
systems \cite{G:2013} and not every system can allow it.
Moreover, our conditions to have 
a bound state can be derived using only the controllable parameters 
of the system, e.g. 
the lattice geometry,  masses and dispersion relations of particles and 
two-body potentials (see examples in Section \ref{sec:examples})
and we do not need to know 
the exact or numerical solutions of two-particle problems.
Recall that 
because of the controllability of collision properties of 
ultracold atoms,  a stable repulsive bound pair 
of ${}_{}^{87}{\rm Rb}$ atoms has been  experimentally observed 
in the optical lattice $\Z^3$ \cite{Wink:2006};
 ultracold heteronuclear molecules 
assembled from fermionic ${}_{}^{40}{\rm K}$ and 
bosonic ${}_{}^{87}{\rm Rb}$  
atoms in $\Z^3$ have been   
produced at a heteronuclear 
Feshbach resonance on both the attractive and 
the repulsive sides of the resonance 
\cite{Osp:2006}; see also \cite{Heo:2012,Zir:2008}.

The present work is organized as follows. 
In Section \ref{sec:aniqla_dsho} we
introduce the notation, the $N$-particle Hamiltonian $\bH$  
and decompose it  into the direct integral of  
Schr\"odinger operators $H(K).$ Cluster operators 
and their spectra are studied in Section \ref{sec:cluster_oper}. 
In Section \ref{sec:HVZ_teoremasi}
we prove the HVZ theorem. Section \ref{sec:bound_states} is devoted to 
 results related to the existence of bound states and
we provide some examples in Section \ref{sec:examples}.
Finally, Section \ref{sec:discussion}
contains some discussions and comments on the  main results.

\section{$N$-particle discrete
Schr\"odinger operator in lattice $\Z^d$}\label{sec:aniqla_dsho}

{\bf Notation.} Let  $N\ge2$ denote the number 
of particles in the system,  $\Z^d$ be a $d\ge1$ dimensional lattice and
$\T^d=\R^d/(2\pi\Z)^d=(-\pi,\pi]^d$  be the $d$-dimensional torus
(the first Brillouin zone, the dual group of $\Z^d$) equipped
with a Haar measure. We abuse the symbol $dq$ to
mean that the integration is upon the variable $q$ over $(\T^d)^m$ 
(or sometimes over its submanifold) for some $m\in\N.$
Elements of $\Z^d$ and $\R^d$ 
will be denoted by $x,y,s,...$ and  
$$
|x|:=|x^{(1)}|+\cdots+|x^{(d)}|,\quad x=(x^{(1)},\ldots,x^{(d)})\in\Z^d\,\,
(\text{or $\R^d$}),
$$
denotes the norm of $x.$
The elements of $\T^d$ are  usually denoted by $p,q,t,\ldots$
and the norm of $p\in\T^d$ is
$$
|p|:=\sqrt{|p^{(1)}|^2+\cdots+|p^{(d)}|^2},\quad p:=(p^{(1)},\ldots,p^{(d)}).
$$
For $p\in\T^d$ and $x\in \Z^d$ we define the duality between $p$
and $x$ as 
$$
p\cdot x:=\sum\limits_{i=1}^d p^{(i)}x^{(i)}.
$$
Given $m\ge1,$  elements of $(\Z^d)^m$ will be denoted 
by bold letters ${\bf x}:=(x_1,\ldots,x_m),$ $x_j\in\Z^d,$ $j=1,\ldots,m.$
The notation is analogous  for $(\T^d)^m.$ 
Moreover, 
$$
|{\bf x}|:=\sum\limits_{j=1}^m |x_j|,\qquad\,\,
|{\bf p}|:=\sqrt{\sum\limits_{j=1}^m |p_j|^2},\qquad\,\,
{\bf p}\cdot {\bf x}:= \sum\limits_{j=1}^m p_j\cdot x_j.
$$

Given $k\in\T^d$ and $m\in\N,$ we also set
\begin{equation}\label{F_K_aniq}
\F_k^m:=\{\q=(q_1,\dots,q_m) \in (\T^d)^m:\,\, q_1+\ldots+q_m=k\}; 
\end{equation}
when $m=N,$ we write shortly $\F_k^N:=\F_k.$
By $\ell^2((\Z^d)^m)$
(resp. $L^2((\T^d)^m)$) we denote the Hilbert space of 
square-summable
(resp. $dk$-square-integrable) functions defined on $(\Z^d)^m$ (resp.
$(\T^d)^m$), $m\ge1.$ For simplicity, set 
$\ell^2((\Z^d)^0)=L^2((\T^d)^0)=\C.$
Elements of $L^2$ is written as $f,g,\ldots$
whereas the notation $\hat f,\hat g,\ldots$ is used for functions on
$\ell^2.$
The symbol $\sigma(A)$ stands for the spectrum of a 
linear operator $A$ and its  essential and discrete spectra are indicated by
$\sigma_\ess(A)$  and $\sigma_\disc(A)$ respectively.  By the standard Fourier transform
we mean the operator $\cF_m:\ell^2((\Z^d)^m)\to L^2((\T^d)^m),$ 
$$
(\cF_m \hat f)({\bf p}):=(2\pi)^{-dm/2} 
\sum\limits_{{\bf x}\in(\Z^d)^m} \hat f({\bf x})e^{\i\,{\bf p}\cdot {\bf x}}. 
$$
Similarly, we introduce the Hilbert space of $\ell^2$ (resp. $L^2$) - functions 
defined on a sublattice  (resp. submanifold) of $(\Z^d)^m$ (resp. 
$(\T^d)^m$). By
$\delta$ we denote  the Dirac-delta function in $\T^d$ concentrated at $0,$
defined  formally as
$$
\delta(q):=(2\pi)^{-d/2}\sum\limits_{x\in\T^d} 
e^{{\rm i}q\cdot x},\qquad q\in\T^d.
$$

\subsection{$N$-particle Hamiltonian}

In coordinate representation, the total Hamiltonian $\widehat \bH,$ 
associated to a
system of $N\ge2$ particles moving in the $d$-dimensional lattice
$\mathbb{Z}^d$ and interacting via  short-range pair potentials 
$\widehat \bV_{ij},$ is defined in 
the Hilbert space $\ell^2((\mathbb{Z}^d)^N)$ as
\begin{equation*} 
 \widehat\bH=\widehat \bH_0 - \widehat \bV, 
\end{equation*}
$$
\widehat \bH_0=\dfrac{1}{m_1}\,\hat \Delta_1\otimes \hat I_d\otimes \cdots
\otimes\hat I_d + \cdots+
\dfrac{1}{m_N}\,\hat I_d \otimes\cdots\otimes \hat I_d \otimes
\hat \Delta_N, 
$$ 
and 
$$
\widehat \bV:=\sum\limits_{1\le i<j\le N} \widehat \bV_{ij}.
$$
Here $m_i\in (0,+\infty]$ -- the mass of particle $i,$  
$\hat I_d$ -- the identity map in $\ell^2(\mathbb{Z}^d),$ 
and $\hat \Delta_i$  -- a
generalized Laplacian -- multidimensional
Laurent-Toeplitz-type operator in $\ell^2(\Z^d):$ 
\begin{equation*}
\hat \Delta_i f(x):= \sum_{s\in {\mathbb{Z}}^d} \hat{\varepsilon}_i(s)
f(x+s),\quad 
f\in\ell^2(\mathbb{Z}^{d}),\quad i=1,\ldots,N.
\end{equation*}
We assume that
\begin{equation}\label{shart_dispersion}
 \begin{cases}
\hat{\varepsilon_i}(y)=\overline{\hat{\varepsilon_i}(-y)},
\qquad y\in \mathbb{Z}^d,\\[2mm]
\exists \gamma>0: \,\,\ \sum\limits_{y\in \Z^d} 
|y|^{\gamma}\,|\hat \epsilon_i(y)|<+\infty,
 \end{cases}
 \qquad i=1,\ldots,N.
\end{equation}

The real-valued continuous function $\varepsilon_i:=\cF_N\hat \varepsilon_i,$
$i=1,\ldots,N,$ is called the {\it dispersion relation of the $i$-th 
normal mode,} associated to the free
particle $i.$  

The pair potential $\widehat \bV_{ij}$
is the multiplication operator by a function
$\hat v_{ij}(x_i-x_j)$
in $\ell^2((\Z^d)^N):$
\begin{equation*}
(\widehat \bV_{ij} f)(x_1,\ldots,x_N) =\hat v_{ij}(x_i
-x_j) \hat f(x_1,\ldots,x_N).
\end{equation*}
We  suppose that 
\begin{equation}\label{shart:potential}
\text{$\hat v_{ij}\in \ell^1(\Z^d)$ 
and is an even function, $1\le i<j\le N$.}  
\end{equation} 
Under assumptions \eqref{shart_dispersion}-\eqref{shart:potential},
the total Hamiltonian $\widehat \bH$ is a bounded  self-adjoint operator
in $\ell^2((\Z^d)^N)$ (see e.g. \cite{Kh:2013} for $N=2$).

In the momentum space $L^2((\T^d)^N)$, $\widehat\bH$ is represented
as
\begin{equation*} 
\bH=\bH_0-\bV, 
\end{equation*}
where
$
  \bH_0=\cF_N\widehat
\bH_0\cF_N^{-1},\quad \bV=\cF_N\widehat \bV\cF_N^{-1}
$
and $\cF_N^{-1}$ is the inverse Fourier transform.

The free Hamiltonian $\bH_0$ is the multiplication operator
$$
 (\bH_0f)(\p)=\cE(\p) f(\p),
$$
by the function  
\begin{equation}\label{obshiy_funskkkaa}
\cE(\p)=\sum\limits_{i=1}^N \frac{1}{m_i}\,\varepsilon_i(p_i), 
\qquad \p:=(p_1,\ldots,p_N)\in (\T^d)^N. 
\end{equation}
The perturbation  $\bV$ is the
sum of partial integral operators $\bV_{ij},$
$i,j=1,\ldots,N,$ $i<j:$
\begin{align*}
(\bV_{ij}f)(\p)=& (2\pi)^{-d/2} \int\limits_{(\T^d)^2}
v_{ij}(p_i-q_i)\delta(p_i+p_j-q_i-q_j)\times
\\&\times f(p_1,\ldots,q_i,\ldots,q_j,\ldots,p_N)\,
dq_idq_j,
\end{align*}
where \textit{the  kernels} $v_{ij}=\cF_1\hat v_{ij},$ 
\begin{equation}\label{potential_kernell}
v_{ij} (p)=(2\pi )^{-d/2}\sum\limits_{x\in \Z^d}  \hat v_{ij}
(x)\,e^{\i p\cdot x},\quad p \in \T^d,
\end{equation}
are real-valued continuous functions on $\T^d.$

\subsection{Decomposition of $\bH$ and representations of fiber operators}
Let $\{\widehat \bU_s\}_{s\in\Z^d}$ be the Abelian group
 of  discrete translations in $\ell^2(({\Z}^d)^N):$ 
$$
(\widehat \bU_s\hat f)(x_1,\ldots,x_N)=\hat f(x_1+s,\ldots,x_N+s),
\quad x_1, \dots, x_2,\, s\in \Z^d.
$$
Via the  Fourier transform $\cF_N$ the family
$\{\widehat \bU_s\}_{s\in \Z^d}$ is unitary-equivalent to the family of
unitary multiplication  operators $\{\bU_s\}_{s\in\Z^d}$ acting in
$L^2((\T^d)^N)$ as
\begin{equation*} 
(\bU_sf)(\p)=\exp(-\mathrm{i}\,(p_1+\cdots+p_N)\cdot s)
f(\p),\quad f \in L^2((\T^d)^N). 
\end{equation*}

Let $\pi_j:(\T^d)^N \to (\T^d)^{N-1},$ $j=1,\ldots,N,$ be 
the projection map defined as
\begin{equation*} 
\pi_j
(p_1,\dots,p_{j-1},p_j,p_{j+1},\dots,p_N)=
(p_1,\dots,p_{j-1},p_{j+1},\dots,p_N),
\end{equation*}
and let $\pi_{jK}$ be the restriction of $\pi_j$ to $\F_K,$
where $\F_K$ is defined in \eqref{F_K_aniq} with $N$ in place of $m.$
As  $\pi_{jK}:\F_K \to (\T^d)^{N-1},$ $K\in\T^d,$ is bijective 
with the inverse mapping given by
\begin{equation*} 
\begin{split}
  (\pi_{jK})^{-1}&(p_1,\dots,p_{j-1},p_{j+1},\dots,p_N)=\\
&\Big(p_1,\dots,p_{j-1}, K-\sum_{i=1,\,i\neq j}^N
p_{i},p_{j+1},\dots,p_N\Big),
\end{split}
\end{equation*}
 $\F_K $ is homeomorphic to  $(\T^d)^{N-1}.$

The decomposition of  the  space $ L^2((\T^d)^N)$  into the direct integral
\begin{equation}\label{dec L2}
L^2((\T^d)^N)=\int_{K\in \T^d}^{\qoshuv} L^2(\F_K)dK 
\end{equation}
yields the corresponding decomposition of the unitary representation
$\bU_s$,
 $s \in \Z^d$, into the  direct integral
\begin{equation*}
\bU_s= \int_{K\in {\T}^d}^{\qoshuv} U_s(K)dK,
\end{equation*}
where
\begin{equation*}
U_s(K)=e^{-\mathrm{i} K\cdot s}I_{L^2(\F_K)}
\end{equation*}
and $I_{L^2(\F_K)}$ is the identity operator on the Hilbert
space $ L^2(\F_K).$  
The  Hamiltonian  $\bH$ obviously
commutes with  $\bU_s$, $s\in \Z^d,$ hence by 
\cite[Theorem XIII.84]{RSIV} the operator
$\bH$ is also decomposed into the von Neumann integral
\begin{equation*} 
\bH= \int_{K \in {\T}^d}^{\qoshuv} \widetilde H(K)dK,
\end{equation*}
associated with the decomposition \eqref{dec L2}.

In the physical literature, the parameter $K\in \T^d$ is called
the {\it $N$-particle  quasi-momen\-tum} and  corresponding
operators $\widetilde H(K),$  $K \in \T^d,$ are called the
{\it fiber operators}.
Observe that given $K\in {\T}^d,$ the fiber 
operator $\widetilde H(K)$ acts in $L^2(\F_K)$ as  
\begin{align}\label{fiber}
 \widetilde H(K)=\widetilde H_0(K)-\widetilde V,
 \end{align}
where 
\begin{equation*} 
\widetilde
H_0(K)f(\p)=\Big(\frac{1}{m_1}\, \epsilon_1(p_1)+\cdots+
\frac{1}{m_N}\,\epsilon_N(p_N)\Big)\,
f(\p),\quad   \p\in \F_K,
 \end{equation*}
and
$$
\widetilde V= \sum_{1\leq i<j\leq N} \widetilde V_{ij}
$$
with
\begin{align}\label{potential_oper_fiber}
(\widetilde V_{ij}f)(\p)=
(2\pi)^{-\frac d2} \int\limits_{\T^d} v_{ij}(t)
f(p_1,\ldots,p_i-t,\ldots,p_j+t,\ldots,p_{N})dt.
\end{align}

Using the unitary operator 
\begin{equation*} 
U_{jK}:L^2(\F_{K}) \rightarrow L^2((\T^d)^{N-1}),\qquad
U_{jK} g=g \circ (\pi_{jK})^{-1},\quad  j=1,\ldots,N, 
\end{equation*}
we define the {\it momentum representation} of  
$\widetilde{H}(K)$  as
$$
H_j(K)=U_{jK}\widetilde{H}(K)U_{jK}^{-1}.
$$
For the simplicity we write $H(K):=H_N(K),$ 
\begin{equation*} 
H(K)=H_0(K)-V,
\end{equation*}
where $H_0(K)$ is the multiplication operator in
$L^2((\T^d)^{N-1})$ by the continuous function
$\cE_K:(\T^d)^{N-1}\to\R$:
$$
\frac{1}{m_1}\,\epsilon_1(p_1)+
\cdots+\frac{1}{m_{N-1}}\,\epsilon_{N-1}(p_{N-1})+
\frac{1}{m_N}\epsilon_{N}(K-p_1-\ldots-p_{N-1}),
$$
and the perturbation $V=\sum\limits_{i<j} V_{ij}$ acts in
$L^2((\T^d)^{N-1})$ with
$$
(V_{ij}f)(p_1,\ldots,p_{N-1})=(2\pi)^{-\frac d2} \int\limits_{\T^d}
v_{ij}(t)
f(p_1,\ldots,p_i-t,\ldots,p_j+t,\ldots,p_{N-1})dt
$$
for $1\le i<j<N,$ and
$$
(V_{iN}f)(p_1,\ldots,p_{N-1})= (2\pi)^{-\frac d2} \int\limits_{\T^d}
v_{iN}(t) f(p_1,\ldots,p_i-t,\ldots,p_{N-1})dt
$$ for $1\le i<N.$

The operator 
\begin{equation*} 
\widehat H(K):=\cF_{N-1}^{-1}H(K)\cF_{N-1} 
\end{equation*}
is called the {\it coordinate representation} of $\widetilde H(K)$ 
in $\ell^2((\Z^d)^{N-1}).$

In what follows any  operator, 
unitarily equivalent to $\widetilde H(K),$ will be called 
the {\it $N$-particle discrete Schr\"odinger operator}.
We use its various representations according to 
their convenience in  applications.

\section{Cluster operators}\label{sec:cluster_oper}

\begin{definition}
A partition $\cC$ of the set $\{1,\ldots,N\}$ into nonintersecting
subsets $C_1,C_2,...,C_\ell$ is called a {\it cluster decomposition}.
Each $C_\nu$ is called a {\it cluster}.
\end{definition}

Given a cluster decomposition $\cC=\{C_1, C_2, \ldots, C_\ell\};$ 
we write $|C_\nu|$ to denote the number of particles in $C_\nu;$
the symbol $ij\in \cC$ means $i,j\in C_\nu$ for some $1\le \nu\le \ell;$ 
analogously, the symbol $ij\notin \cC$ denotes the situation 
in which particles 
$i$ and $j$ are in different clusters
(i.e. $i\in C_{\alpha}$ and $j\in C_{\beta}$ with 
$\alpha\ne \beta$) and $\#\cC$ denotes the number of
elements in $\cC,$ i.e. $\#C=\ell;$ besides, set 
$V^\cC:=\sum\limits_{ij\in \cC} V_{ij},$
$I^\cC:=\sum\limits_{ij \notin \cC} V_{ij}=V-V^\cC.$

\begin{definition}\label{def:cluster_oper}
The operator 
$$
H^\cC(K)=H(K)+I^\cC,\quad K\in \mathbb{T}^d,
$$ 
is called 
the {\it cluster operator} corresponding to a cluster decomposition $\cC.$
We write $\widetilde H^\cC(K)$ (resp. $\widehat H^\cC(K)$) for the 
fiber (resp. coordinate) representation of $H^\cC(K).$
\end{definition}

\subsection{The discrete Schr\"odinger operator corresponding 
to a cluster}\label{subsec:disc_clusters}

Let $C_\nu,$  $1\le \nu\le l,$ be a cluster in a 
decomposition $\cC=\{C_1,\ldots,C_l\}$
and $n_\nu:=|C_\nu|.$ Suppose 
$C_\nu=\{\alpha_1,\ldots,\alpha_{n_\nu}\}\subseteq\{1,\ldots,N\}.$
For $k\in\T^d$ set 
$$
\F_{k}^{n_\nu}: = \{{\bf q}=(q_{\alpha_1},\ldots,q_{\alpha_{n_\nu}})
\in(\T^d)^{n_\nu}:
\quad q_{\alpha_1}+\ldots +q_{\alpha_{n_\nu}}=k\}.
$$
Recall that $\F_k^{n_\nu}$ is homeomorphic to $(\T^d)^{n_\nu-1}.$
In view of \eqref{fiber}  the operator 
$\widetilde h^{C_\nu}(k):L^2(\F_{k}^{n_\nu}) \to
L^2(\F_{k}^{n_\nu}),$ defined as
\begin{equation}\label{lschrop}
\widetilde h^{C_\nu}(k) = \widetilde h_0^{C_\nu}(k)-
\widetilde v^{C_\nu},
\end{equation}
where
$$
(\widetilde h_0^{C_\nu}(k)f) ({\bf p})=
\sum\limits_{\alpha_i\in C_\nu} \frac{1}{m_{\alpha_i}}\,
\epsilon_{\alpha_i}(p_i)\,
f ({\bf p}),\quad {\bf p}\in \F_k^{n_\nu},$$
and 
$$
\widetilde v^{C_\nu} = \sum\limits_{\substack{\alpha_i,
\alpha_j\in C_\nu,\\
\alpha_i<\alpha_j}} \widetilde v_{\alpha_i\alpha_j}
$$
with
$$
(\widetilde v_{\alpha_i\alpha_j}f)({\bf p}) = (2\pi)^{-\frac d2}
\int\limits_{\T^d} v_{\alpha_i\alpha_j}(t)
f(p_{\alpha_1},\ldots,p_{\alpha_i}-t,\ldots,
p_{\alpha_j}+t,
\ldots,p_{\alpha_{n_\nu}}) dt,
$$
is the $n_\nu$-particle discrete Schr\"odinger operator associated to the
Hamiltonian of the particle system $C_\nu.$

\subsection{Spectrum of cluster operators}\label{clust.oper}

Let $K\in\T^d$  and $\cC=\{C_1,\ldots,C_l\},$ $l\ge2.$ 
Set $n_\nu:=|C_\nu|,$ $\nu=1,\ldots,l.$  
It is easy to see that 
$$
\F_K=\bigsqcup\limits_{\substack{k_1,\ldots,k_l\in\T^d,\\
k_1+\cdots+k_l=K}} \F_{k_1}^{n_1}\times\ldots \times
\F_{k_l}^{n_l}:=\bigsqcup\limits_{\substack{k_1,\ldots,k_l\in\T^d,\\
k_1+\cdots+k_l=K}}
 \F(k_1,\ldots,k_l).
 $$
Hence the Hilbert space $L^2(\F_K)$ is decomposed into the von Neumann
direct integral 
\begin{gather}\label{expL2}
L^2(\F_K)=\int_{k_1+\cdots+k_l=K}^{\qoshuv} L^2(\F(
k_1,\ldots,k_l)) d{\bf k},
\end{gather}
where, as stated in the Notation Subsection, 
$d{\bf k}$ denotes the restriction of the Haar measure in $(\T^d)^l$
to the manifold 
$\{{\bf k}=(k_1,\ldots, k_l)\in(\T^d)^l: k_1+\cdots+k_l=K\},$
along which the integration is taken.

Since the fiber cluster operator $\widetilde H^\cC(K)$ commutes with the
decomposable Abelian group of multiplication operators by functions
$\phi_{\bf s}:\F_K\to\C,$ ${\bf s}=(s_1,\dots,s_l)\in(\Z^d)^l,$
$$
\phi_{\bf s}({\bf q})=\exp(\mathrm{i}\sum\limits_{\alpha\in C_1}
q_{\alpha}\cdot s_1)\times \cdots\times
\exp(\mathrm{i}\sum\limits_{\alpha\in C_l} q_{\alpha}\cdot s_l),
$$
the decomposition \eqref{expL2} yields 
the decomposition of  $\widetilde H^\cC(K)$  into the  
direct integral 
\begin{gather}\label{expHCK}
\widetilde H^\cC(K)=\int_{k_1+\cdots+k_l=K}^{\qoshuv} \widetilde
h^\cC(k_1,\ldots,k_l) d{\bf k}.
\end{gather}
The fiber operator $\widetilde h^\cC(k_1,\ldots,k_l),$
$(k_1,\ldots,k_l)\in (\T^d)^l,$ acts in the
Hilbert space
\begin{equation}\label{tensor_L_2}
 L^2(\F(k_1,\ldots,k_l))= L^2(\F_{k_1}^{n_1})\otimes\cdots \otimes
L^2(\F_{k_l}^{n_l})
\end{equation}
as follows:
\begin{align}\label{tenzor}
\widetilde h^\cC(k_1,\ldots,k_l)=& \widetilde h^{C_1}(k_1)\otimes
\widetilde
I^{C_2}(k_2)\otimes\cdots \otimes \widetilde I^{C_l}(k_l)\\
\nonumber
&+\cdots +\widetilde I^{C_1}(k_1)\otimes \widetilde
I^{C_2}(k_2) \otimes \cdots \otimes \widetilde h^{C^l}(k_l),
\end{align}
where $\widetilde I^{C_\nu}(k_\nu)$ is the
identity operator in $L^2(\F_{k_\nu}^{n_\nu})$ and
$\widetilde h^{C_\nu}(k_\nu)$ is the $n_\nu$-particle 
Schr\"odinger operator given by \eqref{lschrop}. We denote 
by $h^{C_\nu}(k_\nu)$ its momentum representation 
acting in $L^2((\T^d)^{n_\nu-1})$
so that $\widetilde h^\cC(k_1,\ldots,k_l)$ is unitarily
equivalent to
\begin{align*}
h^\cC(k_1,\ldots,k_l)=& h^{C_1}(k_1)\otimes
I^{C_2}(k_2)\otimes\cdots \otimes I^{C_l}(k_l)\\
\nonumber
&+\cdots + I^{C_1}(k_1)\otimes 
I^{C_2}(k_2) \otimes \cdots \otimes h^{C^l}(k_l),
\end{align*}
where $k_l:=K-k_1-\ldots-k_{l-1}$ and $I^{C_\nu}(k_\nu)$ is the 
identity operator  in $L^2((\T^d)^{n_\nu-1}).$

\begin{theorem}\label{teo:HDnispektri}
Assume \eqref{shart_dispersion}-\eqref{shart:potential} and let $\cC$
be a cluster decomposition with $\#\cC\ge2.$ 
Then the operator $H^\cC(K)$ has only  essential spectrum and
\begin{align*} 
\sigma(H^\cC(K))=\sigma_\ess(H^\cC(K)) = 
\bigcup\limits_{\substack{k_1,\ldots,k_l \in \T^d,\\
k_1+\ldots+k_l=K} } \sigma(\widetilde
h^\cC(k_1,\ldots,k_l)).
\end{align*}
\end{theorem}

\begin{proof}
We use the fiber representation and
show $\sigma_\disc(\widetilde H^\cC(K))\allowbreak =\emptyset.$ 
Indeed, if $\lambda \in \sigma_\disc(\widetilde H^\cC(K)),$   
by \cite[Theorem XIII.85 (e)]{RSIV}   there would exist  a set 
$A\subset \F(k_1,\ldots,k_l)$ of positive measure such that 
$\lambda \in \sigma_\disc(\widetilde h^\cC(k_1,\ldots,k_l))$ 
for any $(k_1,\ldots,k_l) \in A.$
Let $A=\cup_m A_m$ where $\{A_m\}$ is a pairwise disjoint partition 
of $A$ into sets of positive measure.
For each $(k_1,\ldots,k_l)\in A$ let us choose an associated 
normalized eigenvector
$\psi(k_1,\ldots,k_l)(\cdot)\in L^2(\F(k_1,\ldots,k_l))$ and define 
$$
\Psi_m:= \int_{A_m} \psi(k_1,\ldots,k_l)(\cdot)\,d{\bf k}.
$$
By construction, $\{\Psi_m\}$ is an orthonormal system. Moreover,  
by the definition of the decomposition,
\begin{equation*}
\begin{aligned}
\widetilde H^\cC(K)\Psi_m = & \int_{A_m} 
\widetilde h^\cC(k_1,\ldots,k_l)\psi(k_1,\ldots,k_l)(\cdot)\,d{\bf k}\\ 
= &
\lambda \int_{A_m} \psi(k_1,\ldots,k_l)(\cdot)\,d{\bf k} =\lambda \Psi_m.  
\end{aligned}
\end{equation*}
Hence, $\lambda$ is an eigenvalue of infinite multiplicity, i.e. 
$\lambda \notin \sigma_\disc(\widetilde H^\cC(K)),$ so that
$\sigma(\widetilde H^\cC(K))=\sigma_\ess(\widetilde H^\cC(K)).$

Let us show  now
$$
\sigma(\widetilde H^\cC(K)) =
\bigcup\limits_{\substack{k_1,\ldots,k_l \in \T^d,\\
k_1+\ldots+k_l=K} } \sigma(\widetilde
h^\cC(k_1,\ldots,k_l)).
$$
For shortness, given $\k:=(k_1,\ldots,k_{l-1})\in (\T^d)^{l-1} $ let 
\begin{equation*}
{\bf h}^\cC(\k):=h^{\cC}(k_1,\ldots,k_{l-1}, K-k_1-\ldots-k_{l-1}) 
\end{equation*}
and $U:= \bigcup\limits_{\k} \sigma({\bf h}^\cC(\k)).$
By  \cite[Theorem XIII.85 (d)]{RSIV} 
$
\sigma(H^\cC(K))\subseteq U.
$
On the other hand, if $\lambda\in U,$ there exists 
$\k_0\in (\T^d)^{l-1}$ such that 
$\lambda \in \sigma({\bf h}^\cC(\k_0))$ and   by the norm-continuity 
of $\k\mapsto {\bf h}^\cC(\k)$ there exists a neighborhood 
$O\subset (\T^d)^{l-1}$  of $\k_0$ such that 
$$
(\lambda-\eta,\lambda+\eta)\cap 
\sigma({\bf h}^{\cC}(\k) )\ne\emptyset \qquad \forall \k\in O,\quad \forall 
\eta>0.
$$
Hence, by virtue of \cite[Theorem XIII.85 (d)]{RSIV},  
$\lambda \in \sigma(H^\cC(K)).$ 
Theorem is proved.
\end{proof}

Given cluster decompositions $\cC=\{C_1,\ldots,C_m\}$ 
and $\cD=\{D_1, D_2,\ldots, D_n\},$
we say that $\cC$ is a {\it refinement} of 
$\cD$  if each $D_j$ is a union of some $C_i$'s.

\begin{theorem}\label{teo:spec.clust1}
Assume \eqref{shart_dispersion}-\eqref{shart:potential} 
and let $\cC$ be a refinement of $\cD\ne \cC.$
Then $\sigma(H^\cC(K))\subseteq\sigma_\ess(H^\cD(K)).$
In particular,  $\sigma(H^\cC(K))\subseteq\sigma_\ess(H(K))$
for any cluster decomposition $\cC$ with $\#\cC\ge 2.$
\end{theorem}

\begin{proof} 
We use the coordinate representation.
For simplicity set $\widehat H^{\mathcal L}:=\widehat H^{\mathcal L}(K),$ 
${\mathcal L}=\cC,\cD,$ and
$\widehat H_0:=\widehat H_0(K).$ Note that $\#\cC\ge2$ since $\cC\ne\cD.$
Let $\lambda\in\sigma(\widehat H^\cC)=\sigma_{ess}(\widehat H^\cC).$
By the Weyl criterion there exists a sequence
$\{\hat f_n\}\subset \ell^2((\Z^d)^{N-1})$ weakly converging to $0$ such that
$\|\hat f_n\|=1$ and $\|(\widehat H^\cC-\lambda)\hat f_n\|\to 0$ as $n\to\infty.$
Now by means of $\hat f_n$ we build the sequence
$\{\hat g_r\}\subset \ell^2((\Z^d)^{N-1})$ weakly converging to $0$ such that
$\|\hat g_r\|=1$ and $\|(\widehat H^\cD-\lambda)\hat g_r\|\to 0$ as $r\to\infty.$

Let $\psi\in C^{0,\gamma}([0,\infty))$ (recall that $\gamma$ is given in
\eqref{shart_dispersion}) be 
such that $|\psi(t)|\le1$ and
 $$\psi(t)=
 \begin{cases}
  1,\quad t\ge 2,\\
  0,\quad 0\le t<1.
 \end{cases}
$$
Define 
 $$\rho:(\R^d)^{N-1}\to\R,\quad \rho(y_1,\ldots,y_{N-1})=
 \prod\limits_{\substack{ij\notin C}} \psi(|y_i-y_j|)$$
with $y_N=0\in\R^d.$ 
Since $\psi$ is bounded and H\"older continuous of order
$\gamma,$ so is $\rho.$ Let $c_\rho$ be a H\"older constant of $\rho.$

For $r\in\N$ we define the function
$$
\rho_r:(\Z^d)^{N-1}\to\R,\qquad \rho_r(y)=\rho(y/r).
$$  
Let $\Rho_r$ denote
the multiplication operator by $\rho_r$  in $\ell^2((\Z^d)^{N-1}).$
Observe that $\supp (1-\rho_r)$ is finite,
therefore, the operator $I-\Rho_r$ is compact for any $r\in \N.$ 

Since $\hat f_n$ weakly converges to 0, there exists $N(r)$ 
such that $\|(I-\Rho_r)f_n\|<1/2$
for all $n>N(r).$ This and
the relation $$1=\|\hat f_n\|\le\|(I-\Rho_r)\hat f_n\|+\|\Rho_r\hat 
f_n\|$$ imply that
$\|\Rho_r\hat f_n\|\ge 1/2$ for all $n>N(r).$ We can assume that
$r\mapsto N(r)$ is strictly increasing. 
Now choose a sequence of natural numbers 
$n_1<n_2<\ldots$ such that
$n_r>N(r)$ and consider the sequence  $\hat g_r=\Rho_r\hat 
f_{n_r}/\|\Rho_r\hat f_{n_r}\|$ in $\ell^2((\Z^d)^{N-1}).$
Note that for any $\hat f\in\ell^2((\Z^d)^{N-1})$ we have 
\begin{eqnarray*}
|(\hat g_r,\hat f)|\le &2|(\Rho_r\hat f_{n_r},\hat f)|=2|(\hat f_{n_r},\Rho_r\hat f)|\le\\
&2\|\hat f_{n_r}\|\|\Rho_r\hat f\|=2\|\Rho_r\hat f\|\to0 
\end{eqnarray*}
as $r\to\infty,$ and hence, $g_r$ weakly converges to $0.$

By the definition,
$$
\widehat H^\cD=\widehat H^\cC- 
\sum\limits_{\substack{ij\in \cD,\,ij\notin \cC}} \widehat V_{ij}.
$$
Note that $\widehat V^\cD\Rho_r=\Rho_r\widehat  V^\cD$ as 
$\widehat V_{ij}\Rho_r=\Rho_r\widehat V_{ij}.$ Therefore
\begin{align*}
\|\Rho_r\hat f_{n_r}\| (\widehat H^\cD-\lambda)\hat g_r
=&\, \Rho_r (\widehat H^\cC-\lambda)\hat f_{n_r} +
[\widehat H_0, \Rho_r]\hat f_{n_r}
-\sum\limits_{\substack{ij\in \cD,\,ij\notin \cC}}
\widehat V_{ij}\Rho_r\hat f_{n_r}, 
\end{align*}
where $[A,B]=AB-BA.$
Since $\|\Rho_r\hat f_{n_r}\|\ge 1/2,$ we have
\begin{align} \label{eq1_um}
\|(\widehat H^\cD-\lambda)\hat g_r\|\le &\,
2\|\Rho_r (\widehat H^\cC-\lambda)\hat f_{n_r}\|
+2\|[\widehat H_0, \Rho_r]\hat f_{n_r}\|
+2\sum\limits_{ij\in \cD,  ij\notin \cC}
 \|\widehat V_{ij}\Rho_r\hat f_{n_r}\|.
\end{align}
Observe that 
\begin{equation*} 
\|[\widehat H_0, \Rho_r]\|\le \dfrac{c_\rho}{r^\gamma}\,
\sum\limits_{i=1}^N\left[
\sum\limits_{y\in\Z^d} |y|^\gamma \,|\hat \epsilon_i(y)|\right], 
\end{equation*}
\begin{equation*} 
\|\widehat V_{ij}\Rho_r\|\le
\|\rho\|_\infty \sup\limits_{|y|\ge r}
|\hat v_{ij}(y)|,\quad  \forall ij\in D,\quad ij\notin C, 
\end{equation*}
and 
\begin{equation}\label{eq4_cluster}
\|\Rho_r (\widehat H^\cC-\lambda)\hat f_{n_r}\|\le
\|\rho_r\|_\infty \|(\widehat H^\cC-\lambda)\hat f_{n_r}\|.
\end{equation}
Now the choice $\{\hat f_n\},$  assumptions 
\eqref{shart_dispersion}-\eqref{shart:potential}
and inequalities \eqref{eq1_um}-\eqref{eq4_cluster} imply
$$
\lim\limits_{r\to\infty}\|(\widehat H^\cD-\lambda)\hat g_r\|=0.
$$
Since $\hat g_r$ weakly converges to $0,$ by the
Weyl criterion, $\lambda\in\sigma_\ess(\widehat H^\cD).$
\end{proof}

\section{HVZ theorem for $H(K)$}\label{sec:HVZ_teoremasi}

The main result of this section is the following analogue
of the HVZ theorem.

\begin{theorem}\label{teo:HVZ0}
Assume \eqref{shart_dispersion}-\eqref{shart:potential}.
For any $K\in\T^d$ the essential spectrum of  $H(K)$ 
is an at most countable union of disjoint closed segments;
more precisely, it
consists of the union of the spectra of all two-cluster operators:
\begin{equation*} 
\sigma_{\ess}(H(K)) =
\bigcup\limits_{\cD\in \Xi,\, \#\cD=2} \sigma(H^\cD(K)), 
\end{equation*}
where
$\Xi$ is the set of all cluster decompositions. 
Moreover, $\sigma_\disc(H(K))$ can accumulate only at the edges of 
the constituent segments.
\end{theorem}

Recall that analogous statements for  essential spectra 
of discrete Schr\"odinger operators
have been shown, for instance, in \cite{ALM:07:Nach} 
($N=3:$ proven studying the two-particle channels) and
in \cite{ALA:2003,Mum} ($N=4:$ proven using 
Faddeev-Yakubovskiy equations).
We prove Theorem \ref{teo:HVZ0} using  diagrammatic techniques 
of Hunziker \cite{Hun,RSIV}.

\begin{proof}[Proof of Theorem \ref{teo:HVZ0}] 
The first assertion follows directly from the spectral theory 
of  self-adjoint operators,
as the family of  spectral projections associated 
with $H(K)$ is increasing and $\sigma_\ess(H(K))$  is 
its continuity points \cite{RSI}.

We may suppose that at least one $V_{ij}$ is non-zero, otherwise the 
result is trivial. 
By virtue of Theorem \ref{teo:spec.clust1},
$$
\bigcup\limits_{\cD\in \Xi,\, \#\cD\ge2} \sigma(H^\cD(K))=
\bigcup\limits_{\cD\in \Xi,\, \#\cD=2} \sigma(H^\cD(K))\subseteq 
\sigma_{\ess}(H(K)).
$$

Under the terminology and notation of \cite{Hun}, 
the ``difficult'' part: 
\begin{equation*} 
\sigma_{\ess}(H(K)) \subseteq
\bigcup\limits_{\#\cD\ge2} \sigma(H^\cD) 
\end{equation*}
and the final assertion of the theorem
are proved essentially the same as in \cite{Hun} provided that the 
terms of the expansion
\begin{equation}\label{Wain}
 G(z)=\sum\limits_{n=0}^{\infty} \sum\limits_{(i_1j_1),\ldots,(i_nj_n)}
 G_0(z)V_{i_1j_1}G_0(z)V_{i_2j_2}\ldots G_0(z)V_{i_nj_n}G_0(z),
\end{equation}
corresponding to   connected graphs, are compact, here 
$G_0(z)=(H_0(K)-z)^{-1},$ $G(z)=(H(K)-z)^{-1},$ and $|z|$ is large
enough that the series \eqref{Wain} converges absolutely.
For the convenience of the reader we prove this fact in two steps.

\begin{claim}\label{step:compactgraph}
Let $u\in C^0((\T^d)^{N-1})$ be a non identically zero function 
with Fourier coefficients $\hat u\in \ell^1((\Z^d)^{N-1}).$ 
Assume that $E_0$ is the
multiplication operator by 
the function $u(\cdot)$ in $L^2((\T^d)^{N-1})$ and 
let $T=V_{\alpha_1\beta_1}E_0\ldots E_0V_{\alpha_n\beta_n},$
$\alpha_j,\beta_j\in \{1,\ldots,N\},$ $\alpha_j<\beta_j.$ 
If the graph corresponding to this operator is connected, 
then $T$ is compact.
\end{claim}

It suffices to show that 
$\widehat T(B)$ is relatively compact in $\ell^2((\Z^d)^{N-1}),$
where $B$ is the unit ball in $\ell^2((\Z^d)^{N-1})$ and
$$\widehat T = \cF_{N-1}^{-1} T\cF_{N-1}:=
\widehat V_{\alpha_1\beta_1}\widehat E_0\ldots
\widehat E_0\widehat V_{\alpha_n\beta_n}$$ with
$\widehat E_0=\cF_{N-1}^{-1} E_0 \cF_{N-1}.$
By the Kolmogorov criterion, we need to prove that 
for every $\eta>0$ there exists $R_\eta>0$ such that 
\begin{equation}\label{eq:compactaa}
{\mathcal Q}_R(\hat f):=\sum\limits_{|\x|>R} 
|(\widehat T\hat f)(\x)|^2\le \eta,\qquad R\ge R_\eta, \quad 
\hat f\in B.
\end{equation}

We recall that $\widehat V_{ij},$ $1\le i<j\le N,$ is the multiplication 
operator by  
$\hat v_{ij}(x_i-x_j)$ in $\ell^2((\Z^d)^{N-1})$ with the convention $x_N:=0.$
Since $\|\hat f\|\le1,$ from definitions of $\widehat E_0$ and $\widehat V_{ij}$ 
one can easily verify that 
\begin{align*}
\cQ_R(\hat f)\le &\sup\limits_{|\x|>R} \sum\limits_{\y_1,\ldots,\y_{n-1} 
\in (\Z^d)^{N-1} }
\prod\limits_{j=1}^{n-1} |\hat u(\y_j)|\times \\
&\times \prod\limits_{j=1}^n 
\Big|\hat v_{\alpha_j\beta_j} \Big(x_{\alpha_j} - x_{\beta_j}  + 
\sum\limits_{i=1}^{j-1} \big((\y_i)_{\alpha_j} - (\y_i)_{\beta_j}\big) 
\Big)\Big|^2. 
\end{align*}
Let $M:=\max\limits_{ij} \sup\limits_{x} |\hat v_{ij}(x)|$
(clearly, $M\in (0,+\infty)$),
and given $m\in\N,$
$$\K_L^m:=\{\y=(\y_1,\ldots,\y_m)\in((\Z^d)^{N-1})^m:\,\, 
|\y_i|<L,\,i=1,\ldots, m\},\quad L>0.$$

By assumption on $\hat u,$
$
\sum\limits_{\y_1,\ldots,\y_{n-1} }
\prod\limits_{j=1}^{n-1} |\hat u(\y_{j})|\le \|\hat u\|_{\ell^1}^n<+\infty,
$ 
and thus, given $\eta>0$ there  exists $L_\eta>0$ such that  
\begin{gather*}
\sum\limits_{\y \in ( (\Z^d)^{N-1} )^{n-1} \setminus \K_L^{n-1}}
\prod\limits_{j=1}^{n-1}|\hat u(\y_j)| \le \dfrac{\eta}{2M^{2n}},\qquad 
L>L_\eta. 
\end{gather*}
In particular, for such $L,$
\begin{align}\label{qwert}
\sum\limits_{\y \in ((\Z^d)^{N-1})^{n-1}\setminus \K_L^{n-1}}&
\prod\limits_{j=1}^{n-1}|\hat u(\y_j)|
\prod\limits_{j=1}^n 
\Big|\hat v_{\alpha_j\beta_j} \Big(x_{\alpha_j} - x_{\beta_j}  + 
\sum\limits_{i=1}^{j-1} \big((\y_i)_{\alpha_j} - (\y_i)_{\beta_j}\big) 
\Big)\Big|^2 \le \dfrac{\eta}{2}.
\end{align}
Now let us estimate the finite sum
\begin{align*}
A:= \sum\limits_{\y\in \K_L^{n-1}}&
\prod\limits_{j=1}^{n-1}|\hat u(\y_j)| 
\prod\limits_{j=1}^n 
\Big|\hat v_{\alpha_j\beta_j} \Big(x_{\alpha_j} - x_{\beta_j}  + 
\sum\limits_{i=1}^{j-1} \big((\y_i)_{\alpha_j} - (\y_i)_{\beta_j}\big) 
\Big)\Big|^2. 
\end{align*}
By \eqref{shart:potential} there exists $r_\eta>0$ such that
for any $r>r_\eta,$
$$
\sup\limits_{|y|>r} 
|\hat v_{\alpha_{j}\beta_j}(y)|^2<
\dfrac{\eta}{2\|\hat u\|_{\ell^1}^n M^{2n-2}},\qquad j=1,\ldots,n.
$$ 
Since the graph corresponding to $T$ is connected, 
if $|\x|\to \infty,$ then at least one of $|x_{\alpha_j}-x_{\beta_j}|$ 
tends to $\infty.$ Moreover, if $\y \in \K_L^{n-1},$  
for each $\x\in (\Z^d)^{N-1}$ with $|\x|> Nr+nL$ 
there is $j_0=j_0(\x)$ such that 
$\big|x_{\alpha_{j_0}} - x_{\beta_{j_0}}  + 
\sum\limits_{i=1}^{j_0-1} \big((\y_i)_{\alpha_{j_0}} - 
(\y_i)_{\beta_{j_0}}\big)\big| \ge r,$
and hence for such $\x,$
\begin{align*}
&\prod\limits_{j=1}^n 
\Big|\hat v_{\alpha_j\beta_j} \Big(x_{\alpha_j} - x_{\beta_j}  + 
\sum\limits_{i=1}^{j-1} \big((\y_i)_{\alpha_j} -  (\y_i)_{\beta_j}\big) 
\Big)\Big|^2 \\
&\le  \sup\limits_{|y|\ge r} 
\Big|\hat v_{\alpha_{j_0}\beta_{j_0}} \Big(y_{\alpha_{j_0}} - y_{\beta_{j_0}}  + 
\sum\limits_{i=1}^{j_0-1} \big((\y_i)_{\alpha_{j_0}} - 
(\y_i)_{\beta_{j_0}}\big)\Big)\Big|^2\\
&\times  \prod\limits_{j=1,\,j\ne j_0}^n 
\Big |\hat v_{\alpha_j\beta_j} \Big(x_{\alpha_j} - x_{\beta_j}  + 
\sum\limits_{i=1}^{j-1} \big((\y_i)_{\alpha_j} - (\y_i)_{\beta_j}\big) 
\Big)\Big|^2 
\le \frac{\eta}{2\|\hat u\|_{\ell^1}^n},
\end{align*}
whence, $A\le \eta/2.$  Now \eqref{eq:compactaa} 
follows from this and \eqref{qwert} with the choice 
$R_\eta:=N r_\eta+ nL_\eta.$

\begin{claim}\label{Step2}
The operator $T= G_0(z)V_{\alpha_1\beta_1}G_0(z)\ldots 
G_0(z)V_{\alpha_n\beta_n},$ $1\le \alpha_j<\beta_j\le N,$
$j=1,\ldots,n,$ is compact if and only if 
the corresponding graph is connected. 
\end{claim}

Assume that the corresponding graph is not connected and 
the associated cluster decomposition is
$\cD=\{D_1,\ldots,D_l\}$ with $l\ge2.$ Without loss of ge\-ne\-rality 
we may assume that $N\in D_l.$ Define the Abelian group 
of unitary operators
$U_s,$ $s\in\Z^d$ in $L^2((\T^d)^{N-1})$ as follows:
$$
(U_sf)(\p)= \exp(\i \sum\limits_{\alpha\in D_1} p_\alpha\cdot s ) f(\p).
$$
Clearly, $T$ commutes with $U_s$ as  $V_{\alpha_j\beta_j},$ 
$\alpha_j\beta_j\in \cD,$ $j=1,\ldots,n,$ and
$G_0(z)$ commute with $U_s.$ Choose 
$f_0\in L^2((\T^d)^{N-1})$ such that $Tf_0\ne0.$
Since $U_s\to 0$ weakly as $s\to\infty,$
$U_sf_0\rightharpoonup 0$ in $L^2((\T^d)^{N-1})$ as $s\to\infty,$ 
but 
$$
\|T(U_sf_0)\|=\|U_s(Tf_0)\|=\|Tf_0\|\not \to 0.
$$ 
Hence, $T$ is not compact.

Now assume that the graph is connected. Recall that 
$G_0(z)$ is the multiplication operator by (the non-zero function) 
$u=({\cE_K-z})^{-1},$
$\cE_K(p_1,\ldots,p_{N-1}) = \cE(p_1,\ldots,p_{N-1},K-p_1-\ldots-p_{N-1})$
and $\cE$ is defined in \eqref{obshiy_funskkkaa}.
Since $u\in C^0((\T^d)^{N-1}),$  there exists a sequence of non-zero trigonometric
polynomials $u_j:(\T^d)^{N-1}\to\C,$ $j\in \N,$ such that 
$u_j$ converges uniformly to $u$ as $j\to \infty.$ 
Let $E_j$ be the multiplication operator by the function 
$u_j$ in $L^2((\T^d)^{N-1}):$
note that $E_j\to G_0(z)$ in the operator norm as $j\to+\infty.$ 
Since $\hat u_j= \cF_{N-1}^{-1} u_j \in\ell^1((\Z^d)^{N-1}),$ 
by Step \ref{step:compactgraph} the operators
$$
T_j=E_j  V_{\alpha_1\beta_1} E_j\ldots
V_{\alpha_n\beta_n},\qquad j=1,2,\ldots, 
$$ 
are compact.  Now as $G_0(z)$ and $V_{ij}$ are bounded operators
and $n$ is finite, $T_j$ converges to $T$ in the operator norm, 
implying that $T$ is also compact.
\end{proof}

\section{Existence of $N$-particle bound states}\label{sec:bound_states}

Unless otherwise stated, in this section we 
always suppose  
\begin{equation}\label{shart_exists}
\begin{cases}
\text{a) $\hat \epsilon_i,$ $i=1,\ldots,N,$ satisfies \eqref{shart_dispersion},}\\
\text{b) $\hat v_{ij}\in \ell^1(\Z^1)$ is a nonnegative even function,}\\ 
\text{c) $\sum\limits_{x\in\Z^d} \hat v_{ij}(x)>0,$ $1\le i<j\le N,$ }
\end{cases} 
\end{equation}
so that the system of particles is purely attractive.
Our aim is to  provide some sufficient 
conditions  to have  at  least one $N$-particle bound state 
below the lowest edge of the essential spectrum 
$\Sigma:=\Sigma(K)=\inf \sigma_\ess(H(K)),$ $K\in\T^d.$ 
Recall that by Theorems \ref{teo:HDnispektri} and \ref{teo:HVZ0} 
there exists a cluster decomposition  $\cD=\{D_1,D_2\}$ such that 
$
\Sigma=
\min\limits_{k\in\T^d} \min \sigma({\bf h}^\cD(k)),
$ 
where 
\begin{equation}\label{some_cl_ooo}
{\bf h}^{\cD}(k)= h^{D_1}(k)\otimes I^{D_2}(K-k) +
I^{D_1}(k)\otimes h^{D_2}(K-k),\qquad k\in\T^d. 
\end{equation}
Define 
$$
\Sigma(k):=  \min \sigma({\bf h}^\cD(k)).
$$
By the norm-continuity of $k\mapsto {\bf h}^\cD(k),$  the map
$k \mapsto \Sigma(k)$ is uniformly continuous. 
Hence, there exists $k_0\in\T^d$ 
such that $\Sigma(k_0)=\Sigma.$ 

\begin{lemma}\label{lem:eshmatqul}
Let $z_i(k):=\inf\sigma(h^{D_i}(k)),$ $i=1,2.$ Then 
$\Sigma=z_1(k_0) + z_2(K-k_0).$ 
\end{lemma}

\begin{proof}
Let $n_\nu:=|D_\nu|,$ $\nu=1,2.$
For $\Psi(p,q)=\phi(p)\psi(q),$  
$\phi\in L^2((\T^d)^{n_1-1}),$ $\psi\in L^2((\T^d)^{n_2-1}),$ we have 
$$
({\bf h}^{D}(k_0)\Psi,\Psi) = (h^{D_1}(k_0)\phi,\phi)\|\psi\|_{L^2}^2 +
(h^{D_2}(K-k_0)\psi,\psi)\|\phi\|_{L^2}^2.
$$
Now choosing  $\|\phi\|_{L^2}=\|\psi\|_{L^2}=1$ and taking 
infimum over $\phi$ and $\psi$ we deduce 
\begin{align*}
\Sigma\le & \inf\limits_{\Psi}({\bf h}^{D}(k_0)\Psi,\Psi) \le 
\inf\limits_{\phi} (h^{D_1}(k_0)\phi,\phi) +
\inf\limits_{\psi} (h^{D_2}(K-k_0)\psi,\psi)\\
 = & z_1(k_0) + z_2(K-k_0). 
\end{align*}
On the other hand, for any $\eta>0$ let us choose 
$
\Phi(p,q)=\sum\limits_{l=1}^M \phi_l(p)\psi_l(q)\in
 L^2((\T^d)^{n_1-1})\otimes L^2((\T^d)^{n_2-1})
$
such that both $\{\phi_l\}$ and $\{\psi_l\}$ are orthonormal systems, and
$\Sigma+\eta > ({\bf h}^{D}(k_0)\Phi,\Phi)\|\Phi\|_{L^2}^{-2}.$
But as $\|\Phi\|_{L^2}^2 =M,$ and $(h^{D_1}(k_0)\phi_l,\phi_l)\ge z_1(k_0),$
$(h^{D_2}(K-k_0)\psi_l,\psi_l)\ge z_2(K-k_0),$ we have
\begin{align*}
\Sigma+\eta > &\frac{1}{M}\sum\limits_{l=1}^M
\Big((h^{D_1}(k_0)\phi_l,\phi_l) \|\psi_l\|_{L^2}^2+
(h^{D_2}(K-k_0)\psi_l,\psi_l) \|\phi_l\|_{L^2}^2\Big)\\
\ge & (z_1(k_0) +z_2(K-k_0)) \,\, \frac{\sum\limits_{l=1}^M 
\|\phi_l\|_{L^2}^2 \|\psi_l\|_{L^2}^2 }{M}=z_1(k_0) +z_2(K-k_0).
\end{align*}
Since $\eta>0$ is arbitrary, the assertion of the lemma follows.
\end{proof}

An essential tool in the proof of the 
existence of bound states is the following 

\begin{theorem}\label{teo:existence}
Suppose \eqref{shart_exists}, $K\in\T^d$
and $\Sigma:=\Sigma(K) = \inf\sigma_\ess(H(K)).$
Let $\cD=(D_1,D_2)$ be a cluster decomposition 
such that the cluster operator ${\bf h}^\cD(k),$
defined as in \eqref{some_cl_ooo}, satisfies
$$
\Sigma = \min\limits_{k\in\T^d} \min \sigma({\bf h}^\cD(k))
$$
and let $z_1(k_0)$ and $z_2(K-k_0),$ given in Lemma \ref{lem:eshmatqul}, 
be isolated  eigenvalues of $h^{D_1}(k_0)$ and $h^{D_2}(K-k_0),$ 
respectively. Then $\sigma_\disc(H(K))\cap (-\infty,\Sigma)$ is nonempty.
\end{theorem}

\begin{proof}
Without loss of generality assume that 
$D_1=\{1,\dots,n\}$ and $D_2=\{n+1,\ldots, N\}.$ 
By the norm-continuity of $k\mapsto h^{D_1}(k)$ and $k\mapsto h^{D_2}(K-k),$ 
and by the assumption of the theorem, $z_1(k)$ and $z_2(K-k),$ 
$k\in U_\rho(k_0)\subset\T^d$  are still eigenvalues of $h^{D_1}(k)$ and 
$h^{D_2}(K-k)$ respectively,  where $U_\rho(k_0)$ is a sufficiently 
small $\rho$-neighborhood of $k_0.$
Now we work in  spaces $L^2(\F_K),$ $L^2(\F_k^n)$ and 
$L^2(\F_{K-k}^{N-n}).$
Let $\widetilde \phi(k;\cdot)\in L^2(\F_k^n)$ (resp. 
$\widetilde \psi(K-k;\cdot)\in L^2(\F_{K-k}^{N-n})$) be a 
normalized eigenfunction 
of $\widetilde h^{D_1}(k)$ (resp. $\widetilde h^{D_2}(K-k)$) 
associated with $z_1(k)$ (resp. $z_2(K-k)$), $k\in U_\rho(k_0).$ 
Notice that we have also
$$
\widetilde h^{D_1}(k) \overline{\widetilde \phi(k;\cdot)} =
z_1(k)\,\overline{\widetilde \phi(k;\cdot)} \quad \text{and}\quad 
\widetilde h^{D_2}(K-k) \overline{\widetilde \psi(K-k;\cdot)} =
z_2(K-k)\,\overline{\widetilde \psi(K-k;\cdot)}
$$
for any $k\in U_\rho(k_0),$
hence we can suppose that 
both $\phi(k,\cdot)$ and $\psi(k,\cdot)$ are real-valued.
We extend them to $0$ for  $k\in \T^d\setminus U(k_0).$

For a sequence $\rho_l\in (0,\rho)$ converging to $0,$  
define $\Psi_l\in L^2(\F_K),$ $l\in\N,$ as 
\begin{equation}\label{Psi_l}
\Psi_l(p_1,\ldots,p_N):=|U_{\rho_l}(k_0)|^{-1/2}\,
\chi_{U_{\rho_l}(k_0)}^{}\Big(\sum\limits_{l=1}^n p_l\Big)\,
\phi\Big(\sum\limits_{i=1}^n p_i;p_1,\ldots,p_n \Big)\,
\psi\Big(\sum\limits_{i=n+1}^N p_i;p_{n+1},\ldots,p_N\Big), 
\end{equation}
where $\chi_A$ is the characteristic function of a set $A.$
Then by \eqref{expL2} and \eqref{tensor_L_2},
\begin{align*}
(\Psi_l,\Psi_l) = |U_{\rho_l}(k_0)|^{-1}&
\int_{U_{\rho_l}(k_0)} dk \int_{F_k^n} 
\Big|\phi\Big(k;p_1,\ldots,p_n \Big)\Big|^2\,dp\\
& \times \int_{F_{K-k}^{N-n}}\Big|\psi\Big(K-k;p_{n+1},
\ldots,p_N\Big)\Big|^2\,dp=1.
\end{align*}
Moreover, by virtue of  \eqref{expHCK}, \eqref{tenzor} and
\eqref{potential_oper_fiber}, as well as 
the definitions of $\phi$ and $\psi,$ we have
\begin{align*}
(\widetilde H^\cD(K)\Psi_l,\Psi_l) = & 
|U_{\rho_l}(k_0)|^{-1} \int_{U_{\rho_l}(k_0)} (z_1(k) + z_2(K-k))dk 
\end{align*}
and 
\begin{align*}
(\widetilde V_{nN}\Psi_l,\Psi_l) = & |U_{\rho_l}(k_0)|^{-1}
\int_{\T^d} v_{nN}(t)\, dt
\int_{U_{\rho_l}(k_0)} dk \int_{F_k^n} 
\phi\Big(k-t;p_1,\ldots,p_n-t \Big)
\phi\Big(k;p_1,\ldots,p_n \Big) \,dp\\
& \times 
\int_{F_{K-k}^{N-n}}\psi\Big(K-k+t;p_{n+1},\ldots,p_N+t\Big)
\psi\Big(K-k;p_{n+1},\ldots,p_N\Big)\,dp.
\end{align*}
By assumption \eqref{shart_exists}, $\widetilde V_{ij}\ge0$, thus,
\begin{equation}\label{liminfga_hozirlik}
(\widetilde H(K)\Psi_l,\Psi_l) \le (\widetilde H^\cD(K)\Psi_l,\Psi_l)  - 
(\widetilde V_{nN}\Psi_l,\Psi_l). 
\end{equation}
Since the maps $k\mapsto z_1(k)+z_2(K-k),$  
$k\mapsto \phi(k;\cdot)$ and $k\mapsto \psi(K-k;\cdot)$ 
are continuous in $U_{\rho}(k_0),$ 
taking $\liminf$ as $l\to+\infty$ in \eqref{liminfga_hozirlik}, and
using the relations $\Sigma = z_1(k_0) +z_2(K-k_0)$ and
$\phi(k,\cdot)= 0,$  $k\in \T^d\setminus U_\rho(k_0),$
we obtain 
\begin{align*}
\inf\sigma(H(K)) \le \Sigma - 
&\int_{U_\rho(0)} v_{nN}(t)\,dt  \int_{F_{k_0}^n} 
\phi\Big(k_0-t;p_1,\ldots,p_n-t \Big)
\phi\Big(k_0;p_1,\ldots,p_n \Big) \,dp\\
& \times 
\int_{F_{K-k_0}^{N-n}}\psi\Big(K-k_0+t;p_{n+1},\ldots,p_N+t\Big)
\psi\Big(K-k_0;p_{n+1},\ldots,p_N\Big)\,dp.
\end{align*}
Since
$$
v_{nN}(0) = (2\pi)^{-d/2} \sum\limits_{x\in \Z^d} \hat v_{nN}(x) >0
\qquad\text{(recall \eqref{potential_kernell} and \eqref{shart_exists})},
$$
$$
\int_{F_{k_0}^n} 
\phi\Big(k_0;p_1,\ldots,p_n \Big)
 \phi\Big(k_0;p_1,\ldots,p_n \Big) \,dp =1,
$$
and 
$$
\int_{F_{K-k_0}^{N-n}}\psi\Big(K-k_0;p_{n+1},\ldots,p_N\Big)
\psi\Big(K-k_0;p_{n+1},\ldots,p_N\Big)\,dp=1,
$$
by the continuity, possibly taking smaller $\rho,$  
we may suppose that 
$v_{nN}(t)>0,$  
$$
\int_{F_{k_0}^n} 
\phi\Big(k_0-t;p_1,\ldots,p_n-t \Big)
\phi\Big(k_0;p_1,\ldots,p_n \Big)\,dp >0,
$$
and 
$$
\int_{F_{K-k_0}^{N-n}}\psi\Big(K-k_0+t;p_{n+1},\ldots,p_N+t\Big)
\psi\Big(K-k_0;p_{n+1},\ldots,p_N\Big)\,dp>0,
$$ 
for all $t\in U_\rho(0).$
Then
$\inf\sigma(\widetilde H(K)) < \Sigma,$  whence
$\inf\sigma(\widetilde H(K))\in \sigma_\disc(\widetilde H(K)).$
\end{proof}

\begin{remark}\label{rem:shildirtoy}
According to the proof of
Theorem  \ref{teo:existence}, its  assertion  is still valid 
if we replace the hypothesis \eqref{shart_exists} c)
with a weaker assumption 
$$
\sum\limits_{x\in\Z^d} \,\,\sum\limits_{i<j,\,i\in D_1,\,j\in D_2} 
\hat v_{ij}(x)>0.
$$
\end{remark}

\begin{remark}\label{rem:shildirboy}
The proof of Theorem \ref{teo:existence} may fail 
if there are no intercluster interactions, i.e. the hypothesis
\eqref{shart_exists} c) is replaced by
$$
\sum\limits_{x\in\Z^d} \,\,\sum\limits_{i<j,\,i\in D_1,\,j\in D_2} 
\hat v_{ij}(x)=0.
$$
For example, consider an attractive 
three-particle system in $\Z^2$ in  which 
particles $1$ and $2$ interact  via a zero-range pair potential 
and   particle $3$ does not interact with $1$ and $2,$ i.e. 
$\hat v_{12}=\mu\hat \delta_{x0},$ $\mu>0,$ 
and $\hat v_{13}=\hat v_{23} = 0.$ 
It is known that the ground state energy
$z(k),$ $k\in\T^2,$ of the two-particle operator 
${\bf h^{12}}(k),$ acting in $L^2(\T^2)$ as 
in \eqref{haqiqiy_hola} below,
is always an eigenvalue.
Therefore, by Lemma \ref{lem:eshmatqul}  
the lowest edge $\Sigma(K),$ $K\in\T^2,$ 
of the essential 
spectrum of the three-particle discrete Schr\"odinger operator
$H(K)$ is given by
$$
\Sigma(K)=\min\limits_{k\in \T^2} \Big(z(k) + \epsilon_3(K-k)\Big),
$$
where $\epsilon_3$ is the dispersion relation of the third particle.
Furthermore, there exists $k_0\in\T^2$ such that 
$\Sigma(K)=z(k_0) + \epsilon_3(K-k_0),$ so that for the cluster decomposition
$\cC= \{\{1,2\}, \{3\}\}$ all assumptions (but \eqref{shart_exists} c) 
of Theorem \ref{teo:existence} hold.
However, since $V_{13} = V_{23} = 0,$ for the function $\Psi_l,$
defined in \eqref{Psi_l}, one has 
\begin{equation}\label{mucci}
(\widetilde H(K)\Psi_l,\Psi_l)  =  (\widetilde H^D(K)\Psi_l,\Psi_l), 
\end{equation}
and therefore  if we pass to the limit in \eqref{mucci} 
as $l\to+\infty$ we get the trivial inequality 
$$
\inf\sigma (H(K)) \le \Sigma(K),
$$
which does not give any useful information about $\inf\sigma (H(K)).$
\end{remark}

More practical sufficient condition to have at least 
one $N$-particle bound state is given in
the following corollary of Theorem \ref{teo:existence}.

\begin{theorem}\label{teo:existence_suff}
Suppose \eqref{shart_exists} and 
\begin{equation}\label{ikki_zarra_shart}
\sigma_\disc({\bf h^{ij}}(k))
\cap (-\infty,\min\sigma_\ess({\bf h^{ij}}(k)) )\ne\emptyset  
\quad\text{for all $k\in \T^d,$} 
\end{equation}
where 
\begin{equation}\label{haqiqiy_hola}
{\bf h^{ij}}(k)={\bf h_0^{ij}}(k) - {\bf v_{ij}}, 
\end{equation}
${\bf h_0^{ij}}(k)$ is the multiplication operator by 
$\cE_{ij}(k,q):=\frac{1}{m_i}\,\epsilon_i(q)+\frac{1}{m_j}\,\epsilon_j(k-q)$ 
in $L^2(\T^d)$ and 
$\cF_1^{-1}{\bf v_{ij}} \cF_1$ is the multiplication 
operator by $\hat v_{ij}$ in $\ell^2(\Z^d).$
Then $\sigma_\disc(H(K))\cap (-\infty,\Sigma(K))\ne\emptyset$ 
for any $K\in\T^d.$ 
\end{theorem}

\begin{proof}
The assertion of the theorem follows from the following 
\smallskip 

\noindent 
{\bf Claim. } {\it Let  $\fW$ be  an attractive system of
particles moving in $\Z^d$ for which \eqref{shart_exists} holds 
and given a subsystem $\fS\subseteq \fW,$ 
let $H_{\fS}(K),$ $K\in\T^d,$ denote the 
Schr\"odinger operator, associated to the Hamiltonian 
of $\fS.$ Suppose that 
$$
\inf\sigma(H_{\fS}(K))\in\sigma_{\disc}(H_{\fS}(K))
$$
for every $K\in\T^d$ and 
two-particle subsystem $\fS\subseteq\fW.$ Then 
\begin{equation}\label{plaertte}
\inf\sigma(H_{\fU}(K)) \in  \sigma_{\disc}(H_{\fU}(K)) 
\end{equation}
for any $\fU\subseteq \fW.$ }
\smallskip

We prove \eqref{plaertte} by an induction argument on the 
number of particles $\#\fU$ in $\fU.$

If $\#\fU=1,$ i.e. $\fU$ consists of a single particle $\alpha,$ 
by the definition of the direct integral 
\eqref{dec L2} and the fiber operator \eqref{fiber}, 
$H_{\fU}(K)$ is a multiplication operator by 
$\epsilon_\alpha(K)$ in $\C,$ where $\epsilon_\alpha$ is
the dispersion relation of the particle $\alpha.$
Therefore, 
$$
\sigma(H_{\fU}(K)) = \sigma_{\disc}(H_{\fU}(K)) = \{\epsilon_\alpha(K)\}
$$
and \eqref{plaertte} holds. Recall that, 
by assumption, \eqref{plaertte} is true if $\#\fU=2.$ 

Suppose that $\#\fU \ge3,$  and
\begin{equation}\label{induc_assump}
\inf\sigma (H_{\fS}(k))  \in \sigma_\disc (H_{\fS}(k)),\qquad k\in\T^d,  
\end{equation}
for any $\fS\subsetneq \fU.$
We prove that \eqref{induc_assump} holds also for $\fS=\fU.$ 

Indeed, fix $K\in\T^d.$ In view of Subsection \ref{subsec:disc_clusters}
and by Theorem \ref{teo:HVZ0} 
and  Lemma \ref{lem:eshmatqul}, 
there exist  a cluster decomposition 
$\{\fD_1,\fD_2\}$ of $\fU$ and $k_0\in\T^d$ such that 
$$
\inf\sigma_{\ess}(H_\fU(K)) = 
\inf\sigma(H_{\fD_1}(k_0)) + \inf\sigma(H_{\fD_2}(K-k_0)).
$$
By \eqref{induc_assump}, applied with $\fU=\fD_i,$ $i=1,2,$
 $k=k_0$ and $k=K-k_0,$ we have 
$$
\inf\sigma(H_{\fD_1}(k_0)) \in\sigma_{\disc}(H_{\fD_1}(k_0))\qquad\text{and }\qquad
\inf\sigma(H_{\fD_2}(K-k_0))\in\sigma_{\disc}(H_{\fD_2}(K-k_0)),
$$ 
therefore, by Theorem \ref{teo:existence}, 
$\inf\sigma (H_{\fU}(K))\in\sigma_\disc (H_{\fU}(K)).$
The Claim is proved.
\end{proof}

Under the notation  of Theorem \ref{teo:existence_suff},
define the {\it Birman-Schwinger operator}  $\cB_{ij}(k,z),$
associated to ${\bf h^{ij}}(k),$ 
the nonnegative  compact  and self-adjoint operator in $L^2(\T^d)$ as
\begin{equation}\label{b-sh_oper}
\cB_{ij}(k,z):= {\bf v_{ij}}^{1/2} ({\bf h_0^{ij}}(k)-
z)^{-1}{\bf v_{ij}}^{1/2},\quad z<\min_q \cE_{ij}(k,q),\quad 
1\le i<j\le N. 
\end{equation}

The following corollary of Theorem \ref{teo:existence_suff} provides a 
condition, which  implies \eqref{ikki_zarra_shart}, 
and will be used in the following section.

\begin{corollary}\label{cor:birman_existence}
Let 
\begin{equation}\label{BSh_nomr}
\sup\limits_{k\in\T^d} 
\frac{1}{\lim\limits_{z\to \min_q \cE_{ij}(k,q)-0} 
\|\cB_{ij}(k,z)\|}<1,\quad 1\le i<j\le N.  
\end{equation}
Then $\sigma_\disc(H(K))\cap (-\infty,\Sigma(K))\ne\emptyset$ 
for any $K\in\T^d.$ 
\end{corollary}

\begin{proof}
By the celebrated Birman-Schwinger principle \cite{Bir:1966,Sch:1961},
$z<\min_q \cE_{ij}(k,q)$  is an eigenvalue of 
${\bf h^{ij}}(k)$ if and only if $1$ is an eigenvalue 
of $\cB_{ij}(k,z).$ Since $\cB_{ij}(k,\cdot)$ is increasing
and continuous  on $(-\infty, \min_q \cE_{ij}(k,q)),$
and 
$$
\begin{cases}
\lim\limits_{z\to-\infty} \|\cB_{ij}(k,z)\| =0,\\[2mm] 
\liminf\limits_{z\to \min_q \cE_{ij}(k,q)-0} 
\|\cB_{ij}(k,z)\|>1,\quad k\in\T^d,\quad  \text{(by assumption \eqref{BSh_nomr})},
\end{cases}
$$
there exists $z(k)<\min_q \cE_{ij}(k,q)$ such that 
$\|\cB_{ij}(k,z(k))\|=1.$ By compactness of $\cB_{ij}(k,z(k)),$  
$1$ is its eigenvalue, hence by the Birman-Schwinger principle, 
$z(k)$ is an eigenvalue of ${\bf h^{ij}}(k)$ below the essential spectrum.
Now an application of Theorem \ref{teo:existence_suff}
implies the non-emptiness of 
$\sigma_\disc(H(K))\cap (-\infty,\Sigma(K))$ for any $K\in\T^d.$ 
\end{proof}

In the case of a repulsive  system of particles  (i.e. $\hat v_{ij}\le 0$), 
applying Theorem \ref{teo:existence_suff}
to $-H(K),$ we get
 the following analogue of Theorem \ref{teo:existence_suff}.

\begin{theorem}\label{teo:existence_suff_rep}
Suppose that 
\begin{equation*} 
\begin{cases}
\text{$\hat \epsilon_i,$ $i=1,\ldots,N,$ satisfies \eqref{shart_dispersion},}\\
\text{$\hat v_{ij}\in \ell^1(\Z^d)$ is non-positive even function 
and $\sum\limits_{x\in\Z^d} \hat v_{ij}(x)<0,$ $1\le i<j\le N,$ }
\end{cases} 
\end{equation*}
and 
$\sigma_\disc({\bf h^{ij}}(k))
\cap (\max\sigma_\ess({\bf h^{ij}}(k)),+\infty)\ne\emptyset$ 
for all $k\in \T^d,$
where 
$$
{\bf h^{ij}}(k)={\bf h_0^{ij}}(k) - {\bf v_{ij}},
$$
${\bf h_0^{ij}}(k)$ is the multiplication operator by 
$\cE_{ij}(k,q):=\frac{1}{m_i}\,\epsilon_i(q)+\frac{1}{m_j}\,\epsilon_j(k-q)$ 
in $L^2(\T^d)$ and 
$\cF_1^{-1}{\bf v_{ij}} \cF_1$ is the multiplication 
operator by $\hat v_{ij}$ in $\ell^2(\Z^d),$ $1\le i< j\le N.$
Then $\sigma_\disc(H(K))\cap (\Theta(K),+\infty)\ne\emptyset$ 
for any $K\in\T^d,$ where 
$$
\Theta(K):= \sup\sigma_{\ess}(H(K)),\qquad K\in\T^d. 
$$
\end{theorem}

\begin{remark}
Theorem \ref{teo:existence_suff_rep} is not observed 
in the continuous case as in the latter the essential spectrum is  a
half line.
\end{remark}

\begin{remark}
a) The sign definiteness of $\hat v_{ij}$ is essential in our proofs 
since it provides that all perturbations $V_{ij}$ have 
the same sign so that inequality \eqref{liminfga_hozirlik} is valid.
\smallskip

b) The results hold true also in the presence of several 
particles of infinite mass.%
\end{remark}

\section{Some examples}\label{sec:examples}

In this section we provide some sufficient conditions to have 
at least one $N$-particle bound state in {\it purely attractive} 
systems.  By virtue of Corollary \ref{cor:birman_existence}, we need just 
to provide some conditions for dispersion relations $\hat \epsilon_i$ and 
pair potentials $\hat v_{ij}$ which ensure  the validity of \eqref{BSh_nomr}.

\subsection{$N$-particle system with zero-range potentials}\label{subsec:zero_range}

Let ${\bf h^{ij}}^o(k),$ $k\in\T^d,$  be as in Theorem
\ref{teo:existence_suff} with 
$\hat v_{ij} = \mu_{ij}\hat \delta_{x0},$ $\mu_{ij}>0,$ $1\le i<j\le N,$
and let $\cB_{ij}^o(k,z)$ be the Birman-Schwinger operator, 
associated to ${\bf h^{ij}}^o(k)$ defined as in \eqref{b-sh_oper} 
Then for any $k\in\T^d$ and  $z< \min_p\cE_{ij}(k,p),$
$$
\|\cB_{ij}^o(k,z)\|^2 = \frac{\mu_{ij}}{(2\pi)^{d/2}}\,\int_{\T^d} 
\frac{dq}{\cE_{ij}(k,q)-z}.
$$
Set 
\begin{equation*} 
\cE_{ij}^{\rm min}(k):= \min\limits_{q} \cE_{ij}(k,q),\qquad 
\cE_{ij}^{\rm max}(k):= \max\limits_{q} \cE_{ij}(k,q),\qquad k\in\T^d. 
\end{equation*}
Since 
$$
\begin{cases}
\cE_{ij}^{\rm min}(k) \ge \mathtt{m}_o:=
\frac{1}{m_i} \min\limits_{q} \epsilon_i(q) +
\frac{1}{m_j} \min\limits_{q} \epsilon_j(q) \\ 
\cE_{ij}^{\rm max}(k) \le\mathtt{m^o}:=
\frac{1}{m_i} \max\limits_{q} \epsilon_i(q) +
\frac{1}{m_j} \max\limits_{q} \epsilon_j(q) 
\end{cases}
\qquad \text{for all $k\in\T^d$},
$$
we have\footnote{We suppose 
$\frac{(2\pi)^{d/2}}{\mathtt{m^o} - \mathtt{m_o} }=+\infty$
if $\mathfrak{m}_o=\mathfrak{m}^o$ (i.e. when $\cE_{ij}$ is identically 
constant).} 
$$
A_{ij}(k):=\frac{1}{(2\pi)^{d/2}}\,\int_{\T^d} 
\frac{dq}{\cE_{ij}(k,q) - \cE_{ij}^{\rm min}(k)}\in 
\Big[\frac{(2\pi)^{d/2}}{\mathtt{m^o} - \mathtt{m_o} },+\infty\Big],\qquad 
k\in\T^d,
$$
so that
$\sup\limits_{k\in\T^d} A_{ij}(k)^{-1} \le  (2\pi)^{-d/2}
(\mathtt{m^o} - \mathtt{m_o}) < +\infty.$
Therefore, supposing 
\begin{equation}\label{eqeqeqqqqqq}
\mu_{ij}> \sup\limits_{k\in\T^d} A_{ij}(k)^{-1},  
\end{equation}
we get  \eqref{BSh_nomr}.
Now Corollary \ref{cor:birman_existence} implies that the 
corresponding $N$-particle system has a nonempty discrete spectrum below the 
lower edge of its essential spectrum.

\subsection{$N$-particle system with arbitrary potentials}\label{subsec:arbit_poten}

Let ${\bf h^{ij}}(k),$ $k\in\T^d,$ 
be as in Theorem \ref{teo:existence_suff} with 
\begin{equation*} 
\text{$\hat v_{ij} \ge\mu_{ij}\hat\delta_{x0},$ 
and $\mu_{ij}$ satisfies \eqref{eqeqeqqqqqq},
$1\le i<j\le N,$} 
\end{equation*}
and let ${\bf h^{ij}}^o(k)$ be as in Subsection \ref{subsec:zero_range}.
Since
${\bf h^{ij}}(k) \le {\bf h^{ij}}^o(k),$  
by the 
classical Weyl Theorem on the essential spectrum \cite[Section XIII.4]{RSIV}, 
$$
\sigma_{\rm ess}({\bf h^{ij}}^o(k)) = 
\sigma_{\rm ess}({\bf h^{ij}}(k)) = [\cE_{ij}^{\min}(k), \cE_{ij}^{\max}(k)].
$$
Then applying the Birman-Schwinger principle to 
${\bf h^{ij}}^o(k)$ and using \eqref{eqeqeqqqqqq} we get
$$
\inf \sigma({\bf h^{ij}}(k)) \le 
\inf \sigma({\bf h^{ij}}^o(k)) < \inf \sigma_{\rm ess}({\bf h^{ij}}^o(k)) = 
\cE_{ij}^{\min}(k)=\inf \sigma_{\rm ess}({\bf h^{ij}}(k)),\qquad k\in\T^d.
$$
Now by virtue of Theorem \ref{teo:existence_suff},
$\sigma_\disc(H(K))\cap (-\infty,\Sigma(K))\ne \emptyset$ 
for any $K\in\T^d.$ 

When the dispersion functions are of the form
$\epsilon_i = a_i \epsilon,$ $a_i\in\R,$ $i=1,\ldots,N,$ 
where $\epsilon:\T^d\to \R$ is a conditionally negative definite 
function (see e.g. \cite{ALMM:2006,RSIV})  satisfying \eqref{shart_exists},
by \cite[Proofs of Theorems 1 and 2]{LA:2014},
one has $A_{ij}(k) \ge A_{ij}(0)>0$ for all $k\in\T^d,$ 
and thus, for such systems 
the sufficiency assumption \eqref{eqeqeqqqqqq} reads as 
\begin{equation}\label{cond_neg_definit}
\mu_{ij} > A_{ij}(0)^{-1}. 
\end{equation}

\subsection{The standard discrete Laplacian case} \label{subsec:stand_Lapl}

Suppose that $m_i=1,$   $\hat\epsilon_i=\hat\epsilon,$
$i=1,\ldots,N,$ where
$$
\hat \epsilon(x):=
\begin{cases}
d & x=0,\\
-\frac12 & |x|=1,\\
0 & |x|>1,
\end{cases} 
$$
and 
$\hat v_{ij} = \mu\hat\delta_{x0},$ $\mu>0,$ $1\le i<j\le N.$
Then  
$$
\epsilon(p) = \sum\limits_{l=1}^d (1 - \cos p^{(l)})
$$
is conditionally negative definite and $A_{ij}(k):=A(k)$
for all $ij,$ where 
$$
A(k):=\frac{1}{(2\pi)^{d/2}}\,\int_{\T^d} 
\frac{dq}{2\sum\limits_{l=1}^d \cos\frac{k^{(l)}}{2} (1- 
\cos(\frac{k^{(l)}}{2} -q))}\in (0,+\infty]. 
$$  
By \cite{LH:2011} we have 
$$
A(k)=+\infty 
$$
if $d=1,2,$ or if $d\ge3$ and 
at most two coordinates of $k$ is not equal to $\pi,$
and
$$
A(k)\in(0,+\infty)
$$
if $d\ge3$  and at least three coordinates of $k$ is not equal to $\pi.$
Obviously, $A(k)\ge A(0)$ for any $k\in\T^d,$ 
and if $d\ge3,$ condition \eqref{cond_neg_definit} becomes
\begin{equation*} 
\mu > A(0)^{-1}>0.
\end{equation*}
When $d=1,2,$ \eqref{cond_neg_definit} 
always holds since $A(0)=+\infty$. This shows that a bound state  of an attractive
system of $N$-particles moving in $\Z^d,$ $d=1,2,$ 
always exists. In particular, 
from here we recover the existence results of  \cite{LDK:2016}.

\section{Discussion and conclusion}\label{sec:discussion}

One of the interesting problems in the theory of  
Schr\"odinger operators 
is an appearance of bound states. In the discrete case,
this phenomenon is sufficiently
well-understood in (purely attractive or purely 
repulsive) two-particle systems: if $d=1,2,$ a bound state 
always exists as soon as  particles interact, and 
if $d\ge3,$   bound states come out either 
from the threshold resonances
or from the threshold eigenvalues \cite{ALMM:2006}. 
Moreover, in this case the compactness of the pair potential allows
to establish necessary and sufficient conditions for the 
existence or non existence of bound states by means of  Birman-Schwinger 
operators defined in \eqref{b-sh_oper} (see e.g. 
\cite{ALMM:2006,LH:2011} and references therein).

In multi-particle systems, an appearance of bound states  and a
necessary condition for their existence are yet not very
well-understood. As recalled  in  the introduction, 
in three-particle systems, the Efimov  effect provides 
one sufficient condition: if none of three two-particle
subsystems has a bound state below the threshold of the essential 
spectrum, and at least two of them
have threshold resonances, then the three-particle 
system has infinitely many bound states below the threshold 
\cite{ALM:04:Puan,BT:2017,Gr:2014,Sob:1993,Tam:1991,Yaf:1974}, 
see also review \cite{NE:2017} 
and references therein  for the Efimov physics. 
Besides the Efimov effect, in the continuous case some sufficient conditions  
to pair potentials ensuring  the existence of 
at least one $N$-particle bound state
can be found, for example, in  \cite{PMC:1985,Zhis:1960}.
Moreover, in \cite{BMO:2018} 
authors provide a complete description of spectra 
of a family of Schr\"odinger operators,  associated with a system 
of two  fermions and one different particle, moving on $\R^3$
and interacting via zero range pair potentials. In particular,
using   variational techniques efficiently 
in certain range of parameters (e.g. the mass of the different 
particle and the coupling constant of fermions),
they have proved that 
the three-particle discrete spectrum is non empty.   

The existence of a bound state of a system of three identical bosons 
and of a system of two identical fermions and one different
particle, moving in $\Z^d,$ $d=1,2,$ and interacting via zero-range 
pair potentials (of the same sign), has been recently  
established \cite{LDK:2016,LL:2017}. 
In both works the main method is
the use of   Fredholm determinants associated to  certain 
Birman-Schwinger-type operators.

The main contribution of the present work is to show that 
in a purely attractive (resp. purely repulsive) system of particles,
a multi-particle bound state always exists below  (resp. above)  
the essential spectrum  provided that any two-particle 
subsystem  has a bound state below  (resp. above)  its essential spectrum
(Theorems \ref{teo:existence_suff} and \ref{teo:existence_suff_rep}).
Recall that our method heavily relies on the sign definiteness and 
non-vanishing of pair potentials (Remark \ref{rem:shildirboy}). 
We expect the validity of analogous results for an attractive (resp. repulsive) 
system of identical bosonic systems. This 
and some other extensions of results of the current paper 
for  systems with some particle 
symmetries will be presented in the upcoming paper \cite{Kh:2018}.
An analysis of the appearance of bound states of systems with arbitrary
potentials will be addressed in the future investigations.

\subsection*{Acknowledgements} 
The authors thank to Professor G. Dell'Antonio for
useful discussions.


\begin{thebibliography}{99}

\bibitem{ALA:2003} Albeverio, S., Lakaev, S., Abdullaev, J.: 
On the finiteness of the discrete spectrum of four-particle 
lattice Schr\"odinger operators. Rep. Math. Phys. {\bf 51},  43--70 (2003).

\bibitem{ALM:04:Puan} Albeverio, S., Lakaev, S., Muminov, Z.:
Schr\"{o}dinger  operators on lattices. The Efimov effect and discrete spectrum
asymptotics. Ann. Inst. H. Poincar\'e Phys. Theor. {\bf 5}, 743--772   (2004).



\bibitem{ALM:07:Nach}  Albeverio, S.,  Lakaev, S.,  Muminov, Z.:
On the structure of the essential spectrum for the three-particle Schr\"odinger
operators on lattices. Math. Nachr. {\bf 280}, 699--716 (2007).

\bibitem{ALMM:2006}   Albeverio, S., Lakaev, S., Makarov, K., Muminov, Z.:
The threshold effects for the two-particle Hamiltonians on lattices.
Commun. Math. Phys. {\bf 262}, 91--115 (2006).

\bibitem{BT:2017} Basti, G., Teta, A.: Efimov effect for a three-particle 
system with two identical  fermions. Ann. Henri Poincar\'e {\bf 18},
3975--4003 (2017).

\bibitem{BMO:2018} Becker, S., Michelangeli, A., Ottolini, A.: 
Spectral properties of the 2+1 fermionic trimer with contact interactions.
arXiv:1712.10209 [math-ph].


\bibitem{Bir:1966}  Birman, M.:  On the spectrum of singular 
boundary-value problems. Mat. Sb. (N.S.) {\bf55}, 125--174 (1961). English
transl.: Amer. Math. Soc. Transl. {\bf 53}, 23--80 (1966).


\bibitem{ChM:1980}  Chadan, K.,  Martin, A.: 
A sufficient condition for the existence of bound states 
in a potential without spherical symmetry.
J. Phys. Lett. {\bf41}, 205--206 (1980).


\bibitem{BCK:1986} Bolle, D., Chadan, K., Karner, G.:
On a sufficient condition for the existence of $N$-particle 
bound states. J. Phys. A: Math. Gen. {\bf19}, 2337--2344 (1986).

\bibitem{GSch:97}  Graf, G.,  Schenker, D.: 2-magnon scattering in 
the Heisenberg model. Ann. Inst. Henri Poincar\'e, Phys. Th\'eor. {\bf 67},
91--107 (1997).

\bibitem{G:2013} Gridnev, D.: 
Why there is no Efimov effect for 
four bosons and related results on the finiteness of the discrete spectrum.
J. Math. Phys. {\bf54} (2013).

\bibitem{Gr:2014} Gridnev, D.: Three resonating fermions in flatland: 
proof of the super Efimov effect and the exact discrete spectrum asymptotics.
J. Phys. A: Math. Theor. {\bf 47} (2014).

\bibitem{Heo:2012} Heo, M.-S. {\it et al.}: 
Formation of ultracold fermionic NaLi Feshbach molecules.
Phys. Rev. A {\bf 86} (2012).


\bibitem{Hun} Hunziker, W.: On the spectra of Schr\"odinger 
multi-particle Hamiltonians. Helv. Phys. Acta {\bf 39}, 451--462 (1966).

\bibitem{HS:2000}  Hunziker, W., Sigal, I.:
The quantum $N$-body problem.  J. Math. Phys. {\bf 41}, 3448--3510 (2000).


\bibitem{JBC:1998} 
Jaksch, D.,   Bruder, C.,  Cirac, J.,  Gardiner, C., Zoller, P.:
Cold bosonic atoms in optical lattices. Phys. Rev. Lett. {\bf 81}, 
3108--3111 (1998).

\bibitem{Kh:2013} Kholmatov, Sh.: {\it Asymptotics of Eigenvalues 
of Two-particle Schr\"odinger Operators} (in Russian). 
Lambert Academic Publishing,  Saarbr\"ucken, 2013.

\bibitem{Kh:2018} Kholmatov, Sh., Muminov, Z.: 
Existence of bound states of $N$-body systems 
with particle symmetries in lattice. 
In preparation. 


\bibitem{KS:1980} Klaus, M., Simon, B.: Coupling constant thresholds 
in nonrelativistic quantum mechanics. II. Two cluster thresholds 
in $N$-body systems. Comm. Math. Phys. {\bf 78}, 153--168  (1980).



\bibitem{Lak:1993}  Lakaev, S.: The Efimov effect of a system of three
identical quantum lattice particles. Funkcional. Anal.
Prilozhen. {\bf 27}, 15--28,  (1993).  

\bibitem{LA:2014}  Lakaev, S., Alladustov, Sh.: 
Positivity of eigenvalues of the two-particle Schr\"odinger 
operator on a lattice. TMF  {\bf 178}, 390--402 (2014); 
Theoret. and Math. Phys.  {\bf 178}, 336--346 (2014).


\bibitem{LDK:2016} Lakaev, S., Dell'Antonio, G., 
Khalkhuzhaev, A.: Existence of an isolated band in a system of three 
particles in an optical lattice. J. Phys. A: Math. Theor. {\bf 49} 
(2016). 

\bibitem{LH:2011} Lakaev, S., Kholmatov, Sh.:
 Asymptotics of eigenvalues of two-particle
Schr\"odinger operators on lattices with zero range
interaction. J. Phys. A: Math. Theor. {\bf 44} (2011). 





\bibitem{LL:2017} Lakaev, S., Lakaev, Sh.: 
The existence of bound states in a system
of three particles in an optical lattice. 
J. Phys. A: Math. Theor. {\bf 50} (2017).

\bibitem{LSA:2012} Lewenstein, M., Sanpera, A., Ahufinger, A.:
{\it Ultracold Atoms in Optical Lattices.
Simulating Quantum Many-Body Systems.}
Oxford University Press, Oxford, 2012.

\bibitem{Mat} {Mattis, D.:} {The few-body problem
on a lattice.} Rev. Mod. Phys. {\bf 58},
361--379  (1986).


\bibitem{Mog} Mogilner, A.: Hamiltonians in solid-state physics
as multi-particle discrete Schr\"odinger
operators: problems and results.
Adv. in Sov. Math. {\bf 5}, 139--194  (1991).



\bibitem{Mum} Muminov, M.: A Hunziker-Van Winter-Zhislin theorem for
a four-particle lattice Schr\"odinger operator. 
Theor. Math. Phys. {\bf 148}, 1236--1250 (2006).


\bibitem{Mum9} Muminov, M.:  The infiniteness of
the number of eigenvalues in the gap in the essential
spectrum for the three-particle Schr\"odinger operator on a lattice.
Teoret. Mat. Fiz. {\bf 159}, 299--317  (2009).

\bibitem{NE:2017} Naidon, P.,  Endo, S.: Efimov physics: a review. 
Rep. Prog. Phys. {\bf 80} (2017).

\bibitem{Osp:2006} Ospelkaus, C. {\it et al.}:
Ultracold Heteronuclear Molecules in a 3D Optical Lattice.
Phys. Rev. Lett. {\bf97} (2006).

\bibitem{PMC:1985}  Perez, J., Malta, C.,  Coutinho, F.: 
Bound states of $N$-particles: A variational approach.
J. Math. Phys. {\bf 26}, 2262--2267  (1985).


\bibitem{RSI} Reed, M.,  Simon, B.: {\it  Methods of Modern 
Mathematical Physics. Vol. I:
Functional Analysis.} Academic Press, New York, 1972.

\bibitem{RSIV}  Reed, M.,  Simon, B.:  {\it Methods of Modern Mathematical 
Physics. Vol. IV:
Analysis of Operators.} Academic Press, New York, 1978.

\bibitem{Sch:1961}  Schwinger, J.: On the bound states of 
a given potential. Proc. Nat. Acad. Sci. USA {\bf 47}, 122--129  (1961).

\bibitem{Sig:1976}  Sigal, I.: On the discrete spectrum of the Schr\"odinger 
operators of multi-particle systems. Commun. Math. Phys. {\bf 48},
137--154 (1976).

\bibitem{Sim} Simon, B.: Geometric methods in multi-particle 
quantum systems. Commun. Math. Phys. {\bf 55}, 259--274  (1977).

\bibitem{Sob:1993} Sobolev, A.: The Efimov effect. Discrete spectrum 
asymptotics. Commun. Math. Phys. {\bf 156}, 127--168  (1993).

\bibitem{Tam:1991}  Tamura, H.: The Efimov effect of three-body 
Schr\"odinger operator. J. Funct. Anal. {\bf 95}, 433--459  (1991).

\bibitem{Win} Van Winter, C.: Theory of finite systems of
particles. I. Mat.-Fys. Skr. Danske Vid. Selsk {\bf 1}, 1--60 (1960).

\bibitem{Wall:2015} Wall, M.: {\it Quantum Many-Body Physics of Ultracold 
Molecules in Optical Lattices. Models and Simulation Models.} 
Springer Theses, Cham-Heidelberg-New York, 2015.

\bibitem{Wink:2006} Winkler, K. {\it et al.}: Repulsively bound atom 
pairs in an optical lattice. Nature {\bf 441},  853--856  (2006).


\bibitem{Yaf:1974}  Yafaev, D.: On the theory of the discrete 
spectrum of the three-particle Schr\"odinger operator.
Math. USSR-Sb. {\bf 23}, 535--559  (1974).

\bibitem{Yaf:2000}  Yafaev, D.: {\it Scattering Theory: Some Old 
and New Problems.} Lecture Notes in Mathematics  {\bf 1735}. Springer-
Verlag, Berlin, 2000.


\bibitem{Yo:2017} Yoshitomi, K.:
Finiteness of the discrete spectrum in a
three-body system with point interaction. Math. Slovaca. {\bf 67},
1031--1042 (2017).


\bibitem{Zir:2008} 
 Zirbel, J. {\it et al.}: Heteronuclear molecules in an optical 
 dipole trap. Phys. Rev. A {\bf 78} (2008).


\bibitem{Zhis} Zhislin, G.: Investigation of the spectrum of  the
Schr\"odinger operator for a many particle system. 
Tr. Mosk. Mat. Obs. {\bf 9}, 81--120 (1960).

\bibitem{Zhis:1960} Zhislin, G.: A study of the spectrum of the Schr\"odinger
operator for a system of several particles. Tr. Mosk. Mat. Obs.
{\bf 9},  81 -- 120 (1960).


\end{thebibliography}
\end{document}